\documentclass[aps,prb,twocolumn,showpacs]{revtex4-1}
\pdfoutput=1
\usepackage{graphicx}
\usepackage{amsmath}
\usepackage{amsfonts}
\usepackage{amssymb}
\usepackage{wasysym}
\usepackage{bm}
\usepackage{dsfont}
\usepackage[pstarrows]{curve2e}
\usepackage{color}
\usepackage{multirow}
\usepackage{epsfig}
\usepackage{hyperref}
\hypersetup{
    bookmarks=true,         
    unicode=false,          
    pdftoolbar=false,        
    pdfmenubar=false,        
    pdffitwindow=false,     
    pdfstartview={FitH},    
    pdftitle={Spinon-Phonon Interaction in Algebraic Spin Liquids},    
    pdfauthor={M. Serbyn and P. A. Lee},     
    pdfsubject={},   
    pdfcreator={},   
    pdfproducer={}, 
    pdfkeywords={} {} {}, 
    pdfnewwindow=true,      
    colorlinks=true,       
    linkcolor=[rgb]{0.09, 0.09, 0.5},          
    citecolor=[rgb]{0.09, 0.09, 0.5},        
    filecolor=magenta,      
    urlcolor=[rgb]{0.09, 0.09, 0.5}           
}

\def\bea{\begin{eqnarray}}
\def\eea{\end{eqnarray}}
\def\be{\begin{equation}}
\def\ee{\end{equation}}
\def\bpm{\begin{pmatrix}}
\def\epm{\end{pmatrix}}

\def\Im{\mathop{\rm Im}}

\def\tr{\mathop{\rm tr}}

\newcommand{\corr}[1]{\langle #1\rangle}

\newcommand{\om}{\omega}
\newcommand{\p}{\partial}

\newcommand{\bx}{{\hat {\bm x}}}
\newcommand{\by}{{\hat {\bm y}}}
\newcommand{\bk}{{\bm k}}

\newcommand{\bj}{{\bm j}}
\newcommand{\bi}{{\bm i}}
\newcommand{\bS}{{\bm S}}

\newcommand{\bq}{{\bm q}}
\newcommand{\bQ}{{\bm Q}}
\newcommand{\br}{{\bm r}}
\newcommand{\bt}{{\bm t}}
\newcommand{\ba}{{\bm a}}
\newcommand{\bde}{{\bm e}}
\newcommand{\bsigma}{{\bm \sigma}}
\newcommand{\btau}{{\bm \tau}}
\newcommand{\bu}{{\bm u}}

\newcommand{\calC}{{\cal C}}
\newcommand{\calT}{{\cal T}}
\newcommand{\calG}{{\cal G}}
\newcommand{\Rot}{{\cal R}}
\newcommand{\Ref}{{\cal R}}
\newcommand{\Tra}{{T}}

\newcommand{\SO}[2]{\{#1 | #2 \}}
\newcommand{\1}{\mathds{1}}
\newcommand{\I}{{\rm i}}
\newcommand{\Hpint}{{\cal H}_\text{s-ph}}
\newcommand{\Hploc}{{\hat H}_\text{s-ph}}
\newcommand{\Hploco}{{\hat H}_{\text{s-ph}(0)}}
\newcommand{\Hploct}{{\hat H}_{\text{s-ph}{(1)}}}

\newcommand{\piFs}{$\pi$F$\Square$}
\newcommand{\sFs}{sF$\Square$}
\newcommand{\piFk}{$\pi$F$\davidsstar$}
\newcommand{\zFh}{$0$F$\hexagon$}

\newcommand{\Ttr}{T}
\newcommand{\vq}{{{\vec{q}\,}}}
\newcommand{\vk}{{\vec{k}}}
\newcommand{\vf}{v_F}
\newcommand{\vs}{{v_s}}
\newcommand{\hq}{{\hat q}}
\newcommand{\tg}{{\tilde g}}

\newcommand{\En}{E}
\newcommand{\An}{A}
\newcommand{\Bn}{B}

\newcommand{\mr}[1]{\multirow{2}{*}{#1}}

\newcommand{\picPiF}{
\unitlength=0.8mm
\begin{picture}(35,30)
\put(0,3){
\linethickness{1.5pt}
\put(0,0){\line(1,0){20}}
\put(0,0){\line(0,1){20}}
\put(0,20){\line(1,0){20}}
\put(20,20){\line(0,-1){20}}
\put(20,20){\line(1,0){15}}
\put(20,0){\line(1,0){15}}
\put(20,20){\line(0,1){15}}
\put(0,20){\line(0,1){15}}
\linethickness{2.5pt}
\put(10,0){\vector(1,0){4}}
\put(10,20){\vector(1,0){4}}
\put(30,0){\vector(1,0){4}}
\put(30,20){\vector(1,0){4}}
\put(10,0){\vector(1,0){4}}
\put(0,10){\vector(0,1){4}}
\put(0,30){\vector(0,1){4}}
\put(20,10){\vector(0,-1){4}}
\put(20,30){\vector(0,-1){4}}
\put(0,20){\circle*{2}}
\put(0,0){\circle*{2}}
\put(0,20){\circle*{2}}
\put(20,20){\circle*{2}}
\put(20,0){\circle*{2}}
\put(22,2){\Large 2}
\put(-4,2){\Large 1}
\put(22,22){\Large 3}
\put(-4,22){\Large 4}
\linethickness{1.5pt}
\color{red}
\Dline(-5,-3)(26,-3){4}
\Dline(26,-3)(26,26){4}
\Dline(26,26)(-5,26){4}
\Dline(-5,26)(-5,-3){4}
}
\end{picture}
}
\newcommand{\picpiFPSG}{
\unitlength=0.5mm
\begin{picture}(100,30)
\put(0,0){
\linethickness{1.pt}
\put(0,0){\line(1,0){20}}
\put(0,0){\line(0,1){20}}
\put(0,20){\line(1,0){20}}
\put(20,20){\line(0,-1){20}}
\put(20,20){\line(1,0){15}}
\put(20,0){\line(1,0){15}}
\put(20,20){\line(0,1){15}}
\put(0,20){\line(0,1){15}}
\linethickness{1.5pt}
\put(10,0){\vector(1,0){4}}
\put(10,20){\vector(1,0){4}}
\put(30,0){\vector(1,0){4}}
\put(30,20){\vector(1,0){4}}
\put(10,0){\vector(1,0){4}}
\put(0,10){\vector(0,1){4}}
\put(0,30){\vector(0,1){4}}
\put(20,10){\vector(0,-1){4}}
\put(20,30){\vector(0,-1){4}}
\put(0,20){\circle*{2}}
\put(0,0){\circle*{2}}
\put(0,20){\circle*{2}}
\put(20,20){\circle*{2}}
\put(20,0){\circle*{2}}
\put(22,2){\large 2}
\put(-6,2){\large 1}
\put(22,22){\large 3}
\put(-6,22){\large 4}
}
\linethickness{1.5pt}
\put(45,10){\vector(1,0){20}}
\put(50,14){$\Ref_x$}
\put(75,0){
\linethickness{1.pt}
\put(0,0){\line(1,0){20}}
\put(0,0){\line(0,1){20}}
\put(0,20){\line(1,0){20}}
\put(20,20){\line(0,-1){20}}
\put(20,20){\line(1,0){15}}
\put(20,0){\line(1,0){15}}
\put(20,20){\line(0,1){15}}
\put(0,20){\line(0,1){15}}
\linethickness{1.5pt}
\put(10,0){\vector(-1,0){4}}
\put(10,20){\vector(-1,0){4}}
\put(30,0){\vector(-1,0){4}}
\put(30,20){\vector(-1,0){4}}
\put(10,0){\vector(-1,0){4}}
\put(0,10){\vector(0,-1){4}}
\put(0,30){\vector(0,-1){4}}
\put(20,10){\vector(0,1){4}}
\put(20,30){\vector(0,1){4}}
\put(0,20){\circle*{2}}
\put(0,0){\circle*{2}}
\put(0,20){\circle*{2}}
\put(20,20){\circle*{2}}
\put(20,0){\circle*{2}}
\put(22,2){\large 1}
\put(-6,2){\large 2}
\put(22,22){\large 4}
\put(-6,22){\large 3}
}
\end{picture}
}

\begin{document}

\title{Spinon-Phonon Interaction in Algebraic Spin Liquids}
\author{Maksym Serbyn and Patrick A. Lee}
\affiliation{Department of Physics, Massachusetts Institute of
Technology, Cambridge, Massachusetts 02139}
\date{\today}

\begin{abstract}
Motivated by a  search for experimental probes to access the physics of fractionalized excitations called spinons in a spin liquids, we study the interaction of spinons with lattice vibrations. We consider the case of algebraic spin liquid, when spinons have fermionic statistics and a Dirac-like dispersion. We establish the general procedure for deriving spinon-phonon interactions which is based on a symmetry considerations. The procedure is illustrated for four different algebraic spin liquids:  $\pi$-flux and staggered-flux phases on a square lattice, $\pi$-flux phase on a kagome lattice, and zero flux phase on a honeycomb lattice. Although the low energy description is similar for all these phases, different underlying symmetry group leads to a distinct form of spinon-phonon interaction Hamiltonian.  The explicit form of the spinon-phonon interaction is used to estimate the attenuation of ultrasound in an algebraic spin liquid. The prospectives of the sound attenuation as probe of spinons are discussed.
\end{abstract}


\maketitle

\section{Introduction}

Unambiguous experimental identification of a spin liquid,~\cite{Anderson1,*Anderson2} an exotic ground state of a spin system in a dimension larger than one without a magnetic order remains an open question.~\cite{PAL-Science} A number of theoretical scenarios leads to a ground state with charge-neutral excitations, which carry spin-$1/2$ quantum number and have fermionic statistics. These excitations, called spinons, may have a Fermi surface or a Dirac spectrum,~\cite{PAL-Science,Balents10} and are usually strongly coupled to a gauge field. The latter case, so-called Dirac spin liquid phase, will be the primary subject of attention in the present paper. 

Naively, one would expect that the presence of fermionic excitations in the system could be easily tested experimentally. However, measurements of different thermodynamic quantities, such as spin susceptibility or specific heat, often require subtraction and extrapolation to zero temperature, which can be ambiguous. Transport measurements are limited to a heat conductivity due to neutral character of spinons. The heat transport measurements are  difficult to perform at low temperature.  Finally, neutron scattering, potentially a direct probe of spinons,~\cite{Han12,LeeNagaosa12} requires large single crystals which are not always available. The difficulties with experimental detection of spin liquid phases has lead to a number of theoretical proposals. These include but are not limited to 
Raman scattering,~\cite{Wing-HoRaman} 
inelastic X-rays scattering,~\cite{WingHoRIXS} 
Friedel oscillations,~\cite{MrossPRB84} 
electron spin resonance,~\cite{StarykhPRL07}
impurity physics,~\cite{KimJCMP2008,RibeiroLee,Serbyn12}
and
optical conductivity.~\cite{Potter13,Pilon13} 

Coupling between spinons and phonons can open another channel of decay for phonons, and change the attenuation of ultrasound. Thus, the sound attenuation is another potential probe of spinons.~\footnote{Also, optical phonons has been suggested as a possible mean for detection of VBS order in Ref.~\onlinecite{Hermele-k}.} 
The case of spin liquid with spinon Fermi surface has been recently considered by Zhou and one of us in Ref.~\onlinecite{ZhouLee}.  The spinon-phonon interaction in the long wavelength limit was deduced from hydrodynamical arguments.~\cite{Blount,Tsuneto,ZhouLee}  Assuming that electrons stay in equilibrium with lattice (due to presence of impurities) and making canonical transformation to the moving frame one can easily derive interaction Hamiltonian. The same interaction Hamiltonian can be reproduced from microscopical considerations using so-called deformable ions model.~\cite{Rodriguez,Allen}

However, as was pointed out in Ref.~\onlinecite{ZhouLee}, spinons with a Dirac spectrum require a different treatment. Here we address this problem  and consider the contribution of spinons to the attenuation of ultrasound in a Dirac spin liquid.  In contrary to the case of electrons~\cite{Tsuneto,Rodriguez,Allen} or spinons with Fermi surface,~\cite{ZhouLee} it appears that there is no universal form for the interaction of spinons with acoustic phonons in Dirac spin liquid. This is related to the spinor nature of spinons with Dirac dispersion. Similarly to graphene, spinor structure describes the character of wave function on different sublattices, i.e.\ on the microscopic scale. Consequently, one possible route of deducing the form of spinon-phonon interaction would be to find the change of the Hamiltonian of electronic subsystem induced by the long wave modulation of the lattice parameters, starting from a microscopic Hamiltonian. This procedure, although giving explicit values of coupling constants is not universal. It depends on the microscopic implementation of a different spin liquid phase. Moreover, it is difficult to guarantee that one finds all possible terms in the interaction Hamiltonian.  

We adopt a different approach and use symmetry considerations to find a spinon-phonon interaction Hamiltonian. Similar route has been recently used to deduce electron-phonon interaction in graphene.~\cite{ManesPRB07,Basko} There, representations of the lattice symmetry group of the honeycomb lattice on continuous Dirac fields were used to find all possible symmetry allowed electron-phonon couplings.~\cite{Basko} However, in a Dirac spin liquid, also referred to as an algebraic spin liquid, the notion of symmetry group has to be extended to the projective symmetry group.~\cite{Wen-PSG,Wenbook} 

In this paper we generalize the derivation of spinon-phonon interaction Hamiltonian  to the case of projective representation of lattice symmetry group. The procedure is straightforward, and it requires studying the representation of symmetry group on spin-singlet fermionic bilinears. We consider four different realizations of the Dirac spin liquid: $\pi$-flux~\cite{Hermele-piF} and staggered-flux~\cite{Hermele-sF} phases on a square lattice, as well as $\pi$-flux phase on a kagome lattice~\cite{Hermele-k} and a Dirac spin liquid phase on a honeycomb lattice.  Within low energy effective field theory, all these phases can be described in terms of a Dirac excitations, coupled to a gauge field.~\cite{RantnerWenPRL01,RantnerWenPRB02,VafekPRL02,VafekPRB02,Hermele-piF,Hermele-sF,Hermele-k}  Nevertheless, these phases retain the information about their microscopic origin. This information is encoded in the projective representation of spinon operators under the action of the corresponding symmetry group. 

As we will see below, the projective character of the representation of symmetry group has a profound consequences on a spinon-phonon interaction. For all algebraic spin liquid phases considered here except for the Dirac spin liquid on a honeycomb lattice, only the coupling to the density of spinons is allowed by symmetry at the leading order. However, we find that this coupling is screened due to the gauge field. The sound attenuation coming from the next order terms is suppressed by an extra factor and behaves as $T^3$ at low temperatures. In contrast, the Dirac spin liquid on a honeycomb lattice has a sound attenuation $\propto T$. Although the sound attenuation in all cases is suppressed compared to the case of a spin liquid with a Fermi surface, the spinon contribution is still the dominant process at low temperature and experimental observation of the attenuation of ultrasound due to spinons may be possible. 

The paper is organized as follows. Section~\ref{S:DSLMF} introduces the low energy description of Dirac spin liquid and projective symmetry group. We use the $\pi$-flux phase on a square lattice as an example. Next, we discuss interaction of spinons with phonons in Section~\ref{S:sp-ph}. We describe the general procedure of obtaining spinon-phonon interaction Hamiltonian from symmetry considerations. It is later illustrated  in more details for the $\pi$-flux phase on a square lattice. Having found the interaction Hamiltonian, in Sec.~\ref{S:attn} we study the sound attenuation, concentrating on attenuation of longitudinal sound. Finally, in Sec.~\ref{S:discuss} we summarize and discuss our results. Basic facts from the representation theory of finite groups and details on the calculation of sound attenuation are given in Appendix.

\section{Low energy description of algebraic spin liquids \label{S:DSLMF}}
We review the description of algebraic spin liquid fixed point using language of low energy effective field theory.\cite{RantnerWenPRL01,RantnerWenPRB02,VafekPRL02,VafekPRB02,Hermele-piF,Hermele-sF,Hermele-k} This description suits our purposes since we are interested in coupling between acoustic phonons in the low energy limit. In addition, it provides universal framework applicable to a variety of different algebraic spin liquid phases. In short, three ingredients are needed in order to specify low energy field description of a given algebraic spin liquid phase. These are low energy Hamiltonian for continuous fields, gauge group and representation of projective symmetry group specified by its action on fields. Below we review these ingredients for the general case, as well as illustrate them for the $\pi$-flux phase on a square lattice (\piFs\ phase).  We do not list details for the staggered flux  phase on a square lattice~(\sFs), $\pi$-flux on kagome lattice (\piFk), and Dirac spin liquid phase on honeycomb lattice (\zFh). The reader is referred to Refs.~\onlinecite{Hermele-piF,Hermele-sF,Hermele-k}  for more details, specific for these phases. 

\subsection{Effective field theory, gauge group and projective symmetry group}
The starting point is the spin $S=1/2$ model on some (not necessary Bravais) two-dimensional lattice. The symmetries of spin Hamiltonian are assumed to include SU(2) spin rotations, time reversal and the full lattice group of a given lattice. We write Hamiltonian as
\be    \label{Eq:HHeisenberg}
  H=
  \sum_{\corr{\bi\bj}}
   J_{\bi\bj} \bS_\bi \cdot \bS_\bj
   +
   \ldots,
\ee
where bold indices $\bi, \bj$ denote lattice sites and sum goes over nearest neighbor pairs of sites. Ellipses denotes other short-range interaction terms required to stabilize required spin liquid phase. 

In order to get access to phases with no spin order, we use slave-fermion mean field theory. We represent spin operators using spinon operators $f_{\bi,\alpha}$, $\alpha=\uparrow,\downarrow$:
\be \label{Eq:Sviaf}
  \bS_\bi
  =
  \frac12 f^\dagger_{\bi\alpha} \bsigma_{\alpha\beta} f_{\bj\beta}.
\ee
The mapping between Hilbert spaces is exact, provided one imposes a constraint of no double occupancy, $f^\dagger_{\bi\alpha}  f_{\bi\alpha} =1$, where summation over repeating indices is implied. Such representation of spin has SU(2) gauge redundancy. Therefore, mean field decoupling of spin interaction has to include SU(2) gauge field on the links of the lattice. Saddle points of the mean field theory, depending on their structure~\cite{Wen-PSG,Wenbook} may break this SU(2) symmetry down to U(1) or $\mathds{Z}_2$.

We will be mostly interested in phases with U(1) gauge symmetry (the only SU(2)-symmetric algebraic spin liquid phase considered here is the \piFs\ phase~\cite{Wenbook}). General lattice gauge theory Hamiltonian at the saddle point is
\begin{multline}  \label{Eq:Hmeanfield}
  H_\text{U(1)}
  =
 \frac{h}{2} \sum_{\corr{\bi\bj}} e^2_{\bi \bj} - K\sum_\text{plaq.}\cos (\phi_\square)
 \\
 + \sum_{\corr{\bi\bj}}
   [ \chi_{\bj\bi} e^{-\I a_{\bj\bi}} f^\dagger_{\bi\alpha}  f_{\bj\alpha} + \text{H. c.}],
\end{multline}
where we assume that the choice of $\chi_{\bi\bj}$ leads to a Dirac spectrum. The gauge redundancy lead to appearance of $a_{\bi\bj}$, a compact U(1) vector potential living on the bonds of the lattice, and $e_{\bi\bj}$ which is canonically conjugate electric field taking integer values. We also defined $\phi_\square$, living on the dual lattice, as the lattice version of curl of gauge field $a_{\bi\bj}$.  Gauge constraint is $(\mathop{\rm div} e)_\bi + f^\dagger_{\bi\alpha}f_{\bi \alpha}=1$. Although initially vector potential does not have any kinetic terms, these will be generated due to coupling with fermions and are written in~(\ref{Eq:Hmeanfield}) with coefficient $K$. For $K=0$, $h\gg |\chi_{\bi\bj}|$, one recovers spin model from Eq.~(\ref{Eq:Hmeanfield}). Mean field is a good approximation when $K\gg |\chi_{\bi\bj}|\gg h$ and describes the algebraic spin liquid phase. It has been argued that this phase is stable.~\cite{Hermele-piF,Hermele-sF}  Monopoles are irrelevant, leading to the gauge group becoming non-compact in the low energy limit. Therefore it should be accessible starting from the spin Hamiltonian in Eq.~(\ref{Eq:HHeisenberg}) for some range of initial parameters. More detailed discussion of stability of algebraic spin liquid phase is presented in Refs.~\onlinecite{Hermele-piF,Hermele-sF}.

Provided that choice of $\chi_{\bi\bj}$ leads to a Dirac spectrum, we use continuous fermionic fields to write low-energy Hamiltonian. In the momentum space it reads
\begin{equation} \label{Eq:HCont}
  H
  =
  v_F \int\frac{d^2\bk}{(2\pi)^2}\psi^\dagger_{\sigma a \alpha} [(\bk-\ba)\cdot \btau_{\alpha\beta}] \psi_{\sigma a \beta}.
\end{equation}
The indices $\sigma$, $a$ and $\alpha$ in continuous eight-component fermion field $\psi_{\sigma a \alpha}$ label spin, Dirac valley and sublattice respectively. In what follows, we will use three different sets of Pauli matrices acting in different spaces. Pauli matrices acting on the spin indices are denoted as $\{\sigma^x,\sigma^y,\sigma^z\}$. The second index distinguishes between different Dirac points (valleys) and we label Pauli matrices acting in this space as  $\{\mu^x,\mu^y,\mu^z\}$. Finally, Eq.~(\ref{Eq:HCont}) involves Pauli matrices $\{\tau^x,\tau^y,\tau^z\}$  acting on spinor (sublattice) indices. Below we will omit the sign of the tensor product implying e.g.\ $\sigma^y\tau^x = \sigma^y\otimes\1\otimes \tau^x$.

A particular choice of anzatz $\chi_{\bi\bj}$ naively violates some symmetries of the original Hamiltonian. However, representation Eq.~(\ref{Eq:Sviaf}) is invariant under the gauge transformations
\begin{equation}  \label{Eq:gaugetr}
  f_\bi\rightarrow f_\bi e^{i\theta_\bi},
  \qquad
  a_{\bi\bj}\rightarrow a_{\bi\bj}+\theta_\bi-\theta_\bj.
\end{equation}
Consequently, there is a freedom in the choice of the action of different symmetries: one can supplement them with some gauge transformation. Requiring mean field Hamiltonian~(\ref{Eq:Hmeanfield}) to remain invariant we can fix this freedom and show that all symmetries of original spin Hamiltonian remain unbroken. 

The fact that the action of symmetries on fermionic fields is supplemented by gauge transform has a deep consequences. Rather than forming usual representation of lattice symmetry group, fermions are said to realize \emph{projective representation} of lattice symmetry group.~\cite{Wen-PSG,Wenbook} This representation is fully specified by the action of the lattice symmetry group generators on the fermionic fields. Knowing the mapping from the lattice fields $f_{\bi\alpha}$ to $\psi$, one can easily find the action of generators on the continuous fermionic field. Below, we will demonstrate this procedure for the \piFs\  phase. 

\subsection{Example: $\pi$-flux spin liquid on a square lattice \label{SS:piFanzats}}

\begin{figure}
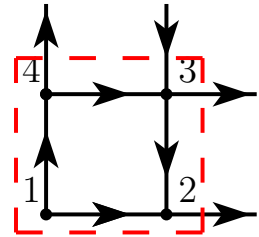

\begin{center}
\picPiF
\caption{ \label{Fig:piF} Choice of the anzats for the \piFs\ phase. Red dashed line encloses the unit cell. Numbers indicate the labeling of sites, used in the main text. Hopping in the direction of arrows is proportional to $+\I$.}
\end{center}
\end{figure}

We use the  the \piFs\ phase~\cite{Hermele-piF} to illustrate the abstract construction presented above. The starting point is the choice specific values of $\chi_{\bi\bj}$ in Eq.~(\ref{Eq:Hmeanfield})  to fix the  mean field anzats.
Our notations are different from those in Ref.~\onlinecite{Hermele-piF}. We quadruple our unit cell, so it includes four sites. Sites within the new unit cell are labeled by an index $m=1\ldots 4$ as in Fig.~\ref{Fig:piF}. The $\chi_{\bi\bj}$ and $a^0_\bi$ are chosen~as
\be
  \chi_{\bi \bi+\bx} = \I,
  \qquad
  \chi_{\bi \bi+\by} = (-)^{\bi_x}\I;
  \qquad
  a^0_\bi
  =
  0.
\ee
Resulting mean field Hamiltonian with the spin index and gauge field omitted is:
\begin{multline} \label{HmffpiF}
  H_\text{mf}
  =
  - \frac12 J \sum_{\br}
  [
  \I(f^\dagger_{\br 2} + f^\dagger_{\br 4} ) f_{\br 1}
  +
  \I (f^\dagger_{\br 3} + f^\dagger_{\br+\ba_2 1} ) f_{\br 4}
  \\
  \I(f^\dagger_{\br+\ba_1 1} - f^\dagger_{\br 3} ) f_{\br 2}
  +
  \I(f^\dagger_{\br+\ba_1 4} - f^\dagger_{\br+\ba_2 2} ) f_{\br 3}
  +
  \text{H. c.}
  ],
\end{multline}
where Bravais lattice vector $\br=n_1 \ba_1+ n_2 \ba_2$ with $n_1$, $n_2$ integers labels unit cells, and  $\ba_1=2\bx$, $\ba_2=2\by$.
Making Fourier transform
\be
  f_{\br m}
  =
  \frac{1}{\sqrt{N}}
  \sum_\bk
  e^{\I \bk\cdot \br}
  f_{\bk m},
\ee
where $N$ is the number of unit cells, we obtain the Hamiltonian in momentum space
\be \label{Eq:HmfpiF}
  H_\text{mf}
  =
  -\frac{J}{2} \sum_\bk
  f^\dagger_{\bk m} H_{mn}(\bk) f_{\bk n},
\ee
where $4\times 4$ matrix $H(\bk)$ is given by
\be  \label{Eq:Hmfkmatrix}
  H_{mn}(\bk)
  =
  \I \bpm
  0       &-1+K_1^* &  0     &  -1 + K_2^* \\
  1-K_1   & 0       &1-K_2^* & 0           \\
  0       &-1+K_2   &  0     & 1-K_1       \\
  1-K_2   & 0       &-1+K_1^*& 0           \\
  \epm,
\ee
with the notation $K_{1,2}=e^{\I \bk\cdot \ba_{1,2}}$. The momentum is measured in the units of the inverse lattice constant of the original lattice (i.e. before enlarging the unit cell), which is set to one in what follows.  We choose the Brillouin zone as $k_x, k_y \in [-\pi/2,\pi/2)$. The energy levels of the Hamiltonian (\ref{Eq:HmfpiF})-(\ref{Eq:Hmfkmatrix}) are doubly degenerate. Therefore there is only one doubly degenerate gapless Fermi point in the Brillouin zone, that is located at the momentum $\bQ=(0,0)$. In the vicinity  of this point the dispersion is described by Dirac fermions and we can define continuum fermion fields. We choose the following representation for two copies of  spinor field $\psi_{a \alpha}(\bk)$:
\begin{subequations}\label{Eq:piFPsi}
\bea
  \psi_{1} (\bk)
  \sim
  \frac{1}{\sqrt{2}}
  \bpm
  \I f_{\bk 2} +   f_{\bk 4}  \\
  -\I f_{\bk 1} -   f_{\bk 3}
  \epm,
  \\
  \psi_{2} (\bk)
  \sim
  \frac{1}{\sqrt{2}}
  \bpm
  f_{\bk 1 } + \I   f_{\bk 3}  \\
  -f_{\bk 2} - \I   f_{\bk 4 }
  \epm,
\eea
\end{subequations}
where the index $a=1,2$ labels different Dirac points, $K_\pm$, both located at the $\Gamma$ point of the Brillouin zone. Consequently, the low-energy continuum Hamiltonian is written as
\begin{equation} \label{Eq:HDiracPiF}
  H_\text{Dirac}
  =
  \vf
  \int \frac{d^2 \bk}{(2\pi)^2}
  \psi^\dagger_{\sigma a}
  (k_x \tau^x+ k_y \tau^y)
  \psi_{\sigma a},
\end{equation}
where made spinor index implicit. Fermi velocity, provided lattice constant is set to one, coincides with $J$, $\vf=J$. 

The continuous fields $\psi_{a\alpha}(\bk)$ realize a projective representation of the lattice symmetry group. 
This representation is fully specified if one knows the action of group  generators on continuous fields. The relevant lattice symmetry group in this case, denoted as $C_{4v}'$,  is different from the point symmetry group of square $C_{4v}$ due to enlarged unit cell. Consequently, the group $C_{4v}'$, in addition to elements from $C_{4v}$, contains translations by a unit vectors of square lattice along $x$ and $y$-axes. The full set of generators, action of which is to be specified below, contains rotation for angle $\pi/2$, $\Rot_{\pi/2}$, reflection of $x$-axis, $\Ref_{x}$~(these are generators of $C_{4v}$) and lattice translation by vector $\ba_1$, $\Tra_{x}$. In addition we also have to specify the action of time-reversal symmetry, $\calT$, and charge conjugation operators, $\calC$.

Let us illustrate the derivation of the action of reflection generator on the continuous fields.  Action of  $\Ref_x$ on the unit cell may be symbolically shown as
\begin{equation} \label{Eq:Rx}
\picpiFPSG
\end{equation}
From here we can understand the action of $\Ref_x$ on the lattice fermion fields, as exchanging fermionic operators with indices $1 \leftrightarrow 2$ and $3 \leftrightarrow 4$. However, in order to leave Hamiltonian~(\ref{HmffpiF}) invariant, this has to be supplemented by gauge transform that changes the sign of all hoppings~[thus reversing the direction of all arrows in the r.h.s.\ of Eq.~(\ref{Eq:Rx})]. One can easily check that transformation   
\begin{subequations}
\begin{eqnarray} \label{Eq:Rxonf}
  f_{\br 1} \rightarrow f_{\br' 2},
  \quad
  f_{\br 2} \rightarrow -f_{\br' 1},
  \\
  f_{\br3} \rightarrow f_{\br'4},
  \quad
  f_{\br4} \rightarrow -f_{\br'3},
  \\
  \br' = \Ref_x \br = (-r_x,r_y)
\end{eqnarray}
\end{subequations}
leaves Hamiltonian~(\ref{HmffpiF}) invariant. The action of $\Ref_x$ on operators $f_{\br n}$ is easy to translate into representation of $\Ref_x$ on the continuous fermionic fields using their definition~(\ref{Eq:piFPsi}). In terms of different sets of Pauli matrices introduced above, it can be written as
\begin{equation} \label{Eq:Rxpsi}
  \psi \to \Ref_x \psi 
  \quad
  \text{with}
  \quad 
  \Ref_x = \I \mu^z \tau^y.
\end{equation}
Using analogous procedure we get the representation of the remaining generators to be 
\begin{eqnarray} \label{Eq:Rpi4}
  \Rot_{\pi/2} &=&\frac12\left(\mu^x+\mu^y)(1+\I \tau^z\right),
  \\ \label{Eq:Tx}
  \Tra_x &=& \mu^y.
\end{eqnarray}
Finally, for the $\pi$-flux phase there exist two additional SU(2) transformations not related to lattice symmetry group. These are time reversal, $\calT$, defined as a antiunitary operator which flips the direction of spin operator~(\ref{Eq:Sviaf}), and charge conjugation, $\calC$. The latter operation may be viewed an $SU(2)$ rotation in the spin space by $\pi$ around $y$-axis,~\cite{Hermele-piF} supplemented by the particle-hole transformation.  On the lattice level (in momentum space) these can be conveniently represented by
\begin{subequations} \label{Eq:TCLattice}
\begin{eqnarray} \label{Eq:TLattice}
  \calT:&\quad
  f_{\bk\sigma1,3}\leftrightarrow f^\dagger_{\bk\sigma1,3}
  ,\qquad
  &
  f_{\bk\sigma2,4}\leftrightarrow -f^\dagger_{\bk\sigma2,4}, 
\\ \label{Eq:CLattice}
  \calC:&\quad
   f_{\bk \uparrow n} \rightarrow   f^\dagger_{-\bk \downarrow n},
  \qquad 
  &
   f_{\bk \downarrow n} \rightarrow -f^\dagger_{-\bk \uparrow n},
\end{eqnarray}
\end{subequations}
where $\calT$-reversal also includes complex conjugation operation, spin indices  were restored. Although these definitions may look counterintuitive, one can check that the charge conjugation~(\ref{Eq:CLattice})  indeed leaves the spin operator invariant, whereas the time-reversal symmetry, defined as in~(\ref{Eq:TLattice}), flips the direction of spin operator in Eq.~(\ref{Eq:Sviaf}). Mapping this action to continuous fermionic fields, we have:
\begin{subequations} \label{Eq:TCPsi}
\begin{eqnarray} \label{Eq:TPsi}
  \calT:&\qquad
  \psi \rightarrow &  \mu^z \tau^z (\psi^\dagger)^T,
\\ \label{Eq:CPsi}
  \calC:&\qquad
  \psi \rightarrow & (\I \sigma^y) (\I \mu^x \tau^x) (\psi^\dagger)^T.
\end{eqnarray}
\end{subequations}

As we pointed out earlier, the representation of the lattice symmetry group on fermions is projective. This can be easily seen from the action of generators, Eqs.~(\ref{Eq:Rxpsi})-(\ref{Eq:Tx}), if one tries to test some group identities. For example, $(\Rot_{\pi/2})^4$ is a trivial transformation. However, using the explicit form of the representation of  $\Rot_{\pi/2}$ for continuous fermionic fields, Eq.~(\ref{Eq:Rpi4}), we find
\begin{equation} \label{Eq:Rpi24}
  (\Rot_{\pi/2})^4  = -\1.
\end{equation}
Thus all group identities hold only up to some gauge transformation, which leaves the Hamiltonian invariant.

\section{Spinon-phonon interaction \label{S:sp-ph}}

Having a low energy description of Dirac spin liquid phases at our disposal, in this section we consider the spinon-phonon interaction.

As we explained in the introduction, the hydrodynamic approach,~\cite{Blount,Tsuneto,ZhouLee} applicable for the case of spin liquid with the Fermi surface, is not straightforward to use in our case. It is the presence of spinor structure, inherently related to the microscopic details such as two inequivalent sublattices, that prevents application of hydrodynamical arguments. Of course one can always resort to the microscopic derivation of spinon-phonon interaction. Despite the advantage of giving specific values of coupling constants, this route is highly non-universal and is not guaranteed to yield all possible couplings. 

We present universal procedure for finding all possible terms in spinon-phonon interaction Hamiltonian, allowed by symmetry. First, we introduce phonons and the general form of the spinon-phonon interaction Hamiltonian. After this the general idea behind the procedure is described. Implementation of this procedure for the \piFs\ phase serves as an example. Finally, we present results for other Dirac spin liquid phases and discuss the underlying physics. The derivation of these results extensively relies on a representation theory for finite groups. Necessary concepts, as well as basic facts about point groups of square, kagome and honeycomb lattices are listed in Appendix~\ref{A:repr}.

\subsection{Spinon-phonon interaction Hamiltonian from symmetry considerations \label{SS:PhononsGeneral}}

We start with specifying conventions for the spinon-phonon interaction Hamiltonian. It is written using the operator $\Hploc(\bk,\bq)$ as 
\begin{equation} \label{Eq:Hsp-phDef}
  \Hpint
  =
  \sum_{\bk,\bq} 
  \psi^\dagger(\bk+\bq)
  \Hploc(\bk,\bq)
  \psi(\bk).
\end{equation}
In what follows we will refer to  the operator $\Hploc(\bk,\bq)$ itself as a spinon-phonon interaction Hamiltonian. Normally, the operator $\Hploc(\bk,\bq)$ obtained from the procedure described above, would contain only zeroth order terms in the distance from the Dirac point, $\bk$, $\Hploc(\bk,\bq)= \Hploco(\bq)$. As we shall see, in some cases, all such terms vanish. Then, to find a non-zero interaction Hamiltonian, we allow the presence of terms, linear in $\bk$, $\Hploct(\bk,\bq)$ and the total Hamiltonian will be written as a sum:
\begin{equation} \label{Eq:Hsp-phExp}
  \Hploc(\bk,\bq)
  =
  \Hploco(\bq)
  +
   \Hploct(\bk,\bq),
\end{equation}
In what follows, we restrict ourselves to the first non-vanishing term in this expansion.  Phonons enter $\Hploc(\bk,\bq)$ via the $\bq$ Fourier component of displacement field, $\bu(t,\br)$.  In the second quantized language the displacement field is written as
\begin{equation} \label{Eq:u(r)}
  \bu(t,\br)
  =
  \sum_{\bq,\mu}
  \sqrt{\frac{\hbar}{2 S \rho\om_\bq}}
  \bde_{\bq \mu}(a_{\bq\mu} e^{-\I \omega_\bq t +\I\bq\cdot\br}+a^\dagger_{-\bq\mu}e^{\I \omega_\bq t -\I\bq\cdot\br}),
\end{equation}
where index $\mu=\text{L,T}$ labels longitudinal and transverse modes of acoustic phonons, and $\bde_{
\bq\mu}$ is the corresponding polarization vector. The dispersion of phonons is assumed in the form $\omega_\bq = \vs |\bq|$, where $\vs$ is the sound velocity. The $\rho$ is defined as a mass density per layer, and $S$ is the area. For simplicity we consider only in-plane phonon modes.

Although we work in continuum limit, it is the lattice symmetry group and its representations, which determines the properties of low energy (acoustic) phonons and spinon excitations. Phonons are described using vector $\bu(t,\br)$, describing displacement at a given point $\br$ due to deformation. As a uniform displacement of the entire lattice, $\bu(t,\br)  = \bu_0$, leaves system invariant, acoustic phonons can couple to spinons only via spatial derivatives of $\bu(\br)$~[we ignore coupling to the time derivative, as it is suppressed by the ratio of sound and Fermi velocities]. Set of all spatial derivatives, $\partial_i u_j (\br)$ or $-\I q_i u_j (\bq)$ in the Fourier space, transforms as a rank-two tensor under lattice symmetry group. Representation of a lattice symmetry group on a rank-two tensor can be split as a sum of  irreducible representations.  Symbolically this is written as
\begin{equation} \label{Eq:E1E1symb}
  E_1\times E_1 
  =
  \sum_j  \oplus D^\text{ph}_j,
\end{equation}
where $E_1$ is vector representation, and $D^\text{ph}_j$ are (possibly repeating) irreducible representations. Acoustic phonon modes can be classified using irreducible components, present in this decomposition. 

\begin{table*}
\begin{center}
\begin{tabular}{l c c c c c c ccc cccc c c}
\hline\hline
Representation   & $A_1$  & $A_2$       &          $\Bn'_1$ &              $\Bn'_2$ &            $E_1$   &             $E_2$    &                               $\En'_1$ &                                 $\En_3$&                                $\En_4$         &                                 $\En_5$ \\
Basis& $\1$ & $\tau^z$ & $\mu^z$ & $\mu^z\tau^z$ &$\tau^x,\tau^y $ & $\mu^x,\mu^y$& $\mu^z\tau^x,\mu^z\tau^y$ & $ \mu^x\tau^z,\mu^y\tau^z$&$\mu^x\tau^y,\mu^y\tau^x$&$\mu^x\tau^x,\mu^y\tau^y$\\
$\calT$-inv      &  $-$   &    $-$      &      $-$       &              $-$   &             $+$   &             $+$     &                             $+$     &                                 $+$  &                               $-$            &                                  $-$   \\
$\calC$-inv      &  $-$   &    $+$     &      $+$      &              $-$   &             $-$    &              $-$     &                             $+$     &                                 $+$  &                                $-$           &                                   $-$   \\
\hline\hline
\end{tabular}
\caption{\label{Tab:C'4vpiF} Explicit form of basis  in terms of tensor products of Pauli matrices for irreducible representations of $C'_{4v}$ contained within $G^{\pi F}_{\psi^\dagger\psi}$. Last two rows show properties of basis elements under time-reversal and charge conjugation. Plus implies invariance, whereas minus indicates a change of sign under the action of corresponding symmetry.}
\end{center}
\end{table*}

Spinons have fermionic statistic, thus minimal coupling to phonons must involve bilinears of $\psi$ field. In contrary to phonons, continuous spinon fields $\psi$  realize \emph{projective} representation of the lattice symmetry group. Action of lattice symmetries on $\psi$ in general includes the gauge transformation, and all identities between generators are valid modulus gauge transformation~[for example, see Eq.~(\ref{Eq:Rpi24})]. Similar to a single field $\psi$, general spinon bilinear also realizes projective representation of lattice symmetry group. However, there exists a \emph{subset} of spinon bilinears which transform under regular representation of symmetry group. For the case when the gauge group is SU(2) these are bilinears which are singlets under SU(2).   Whereas for abelian gauge groups, like U(1) or $\mathds{Z}_2$, \emph{all} bilinears realize regular representation, as gauge component cancels. 

Regular representation of the lattice symmetry group on (a subset of) spinon bilinears can be split into irreducible representations $D^{\psi^\dagger\psi}_j$,
\begin{equation} \label{Eq:Gpsipsisymb}
  G_{\psi^\dagger\psi}
  =
  \sum_j \oplus D^{\psi^\dagger\psi}_j.
\end{equation}
We note, that invariant fermionic bilinears for \piFs, \sFs\ and  \piFk\ phases were identified in Refs.~\onlinecite{Hermele-piF,Hermele-sF,Hermele-k}.  This corresponds to finding all trivial components contained within decomposition~(\ref{Eq:Gpsipsisymb}).

The product of two irreducible representations, $D^{\text{ph}}_i\times D^{\psi^\dagger\psi}_j$ contains a trivial representation within itself if, and only if these representations coincide, $D^{\text{ph}}_i\equiv D^{\psi^\dagger\psi}_j$. As the spinon-phonon interaction Hamiltonian has to be invariant under the action of the symmetry group, we can construct all symmetry allowed couplings by pairing identical irreducible components between splittings~(\ref{Eq:E1E1symb}) and~(\ref{Eq:Gpsipsisymb}). The presence of additional symmetry operations, such as time-reversal or charge conjugation may impose further restrictions on the obtained set. 

\subsection{Example: derivation for the $\pi$-flux phase  \label{SS:PhononsPiF}}

We use the $\pi$F phase as an example for an illustration of the abstract procedure outlined above. For phonons, the underlying symmetry group is $C_{4v}$~(see Appendix~\ref{AA:square}).   Using characters Table~\ref{Tab:C4v}, we find explicit form of the decomposition~(\ref{Eq:E1E1symb}) for the present case:
\begin{equation} \label{Eq:E1E1C4v}
 E_1\times E_1
  =
  A_1\oplus A_2 \oplus  B_1 \oplus B_2.
\end{equation}
$A_1$ here and in what follows always denotes the trivial representation. All other representations are also one-dimensional, thus the action of corresponding group elements can be inferred from Table~\ref{Tab:C4v} in the Appendix. In terms of components of two vectors $(q_x,q_y)$ and $(u_x, u_y)$, the basis functions of these representations are 
\begin{subequations} \label{Eq:E1E1bases}
\begin{eqnarray}
 A_1 : u_{xx} + u_{yy},
 \qquad
 A_2 : u_{xy} - u_{yx},
 \\
 B_1 : u_{xx} - u_{yy},
 \qquad
 B_2 :  u_{xy} + u_{yx},
\end{eqnarray}
\end{subequations}
where we introduced shorthand notation
\begin{equation} \label{Eq:uxx-def}
 u_{ij}\equiv q_i u_j.
\end{equation}

While introducing the ansatz for \piFs\ phase in Section~\ref{SS:piFanzats} we used unit cell consisting of four lattice sites. This allowed us to write relations between continuous fields and microscopic spinon operators in a simple form. However, the price to pay is that spinons now transform under the symmetry group $C_{4v}'$ which is larger than point symmetry group of the square lattice.  In addition to transformations from the point group of the square $C_{4v}$, group $C_{4v}'$ includes lattice translations by unit vector in $\hat x$ and $\hat y$ directions. The details about irreducible representations of  group $C_{4v}'$ are worked out in Appendix~\ref{AA:square}. It has eight one-dimensional irreducible representations labelled as $A_{1,2}, B_{1,2}$ and $A'_{1,2}, B'_{1,2}$  in Table~\ref{Tab:C'4virreps}, and six two-dimensional representations denoted as $E_{1}, E'_{1}, E_{2,\ldots, 5}$. To find the splitting of representation of $C_{4v}'$ on spinon bilinears into irreducible components we use natural basis: all spin singlet bilinears can be enumerated using tensor products of Pauli matrices acting in sublattice and valley space,
\begin{equation} \label{Eq:BasisBilinears}
  \psi^\dagger M \psi, 
  \qquad 
   M \in \{\1, \tau^i, \mu^i, \tau^i \mu^j \}.
\end{equation}
With the help of the characters Table~\ref{Tab:C'4virreps}, the $16$-dimensional representation $G^{\pi F}_{\psi^\dagger\psi}$ is reduced into direct sum of four one-dimensional  and six two-dimensional representations  as 
\begin{multline} \label{Eq:GpiF}
  G^\text{$\pi$F}_{\psi^\dagger \psi}
  =
  A_1\oplus A_2 \oplus \Bn'_1 \oplus \Bn'_2 \oplus E_1 \oplus \En'_1 \oplus E_2 
  \\
  \oplus \En_3 \oplus \En_4 \oplus \En_5. 
\end{multline} 

Using the explicit basis~(\ref{Eq:BasisBilinears}), we can find to what irreducible representation a given matrix belongs. Identity matrix $\1$  corresponds to the trivial representation, $A_1$. Next, one can check that both matrices $\mu^z$, $\tau^z$, and their product, $\mu^z\tau^z$, are invariant (up to a sign) under the action of all generators of $C_{4v}'$, Eqs.~(\ref{Eq:Rxpsi})-(\ref{Eq:Tx}). Therefore the matrices $\mu^z$, $\tau^z$, and $\mu^z\tau^z$  form the basis of one-dimensional representations $A_2$, $\Bn'_1$ and $\Bn'_2$ respectively. Matrices $(\tau^x, \tau^y)$, $(\mu^z\tau^x, \mu^z\tau^y)$, and $(\mu^x, \mu^y)$ constitute basis of two-dimensional irreducible representations $E_1$, $E'_1$ and $E_2$ respectively. Finally, after some algebra, the remaining six matrices from~(\ref{Eq:BasisBilinears}) can be split into pairs that realize the basis for representations $E_{3},\ldots, E_{5}$ as shown in the Table~\ref{Tab:C'4vpiF}. The last two rows in Table~\ref{Tab:C'4vpiF} display the symmetry of corresponding matrices under the action of time reversal and charge conjugation operations. 

Explicit  decompositions, Eqs.~(\ref{Eq:E1E1C4v}) and (\ref{Eq:GpiF}) give us allowed couplings between spinons and phonons.  Only identical irreducible representations can be coupled between themselves. Comparing Eqs.~(\ref{Eq:E1E1C4v}) and (\ref{Eq:GpiF}) we see that only two first terms in both direct sums coincide.  Thus one may expect the allowed couplings to be described by contraction between  $A_1$ ($A_2$) components in different sums. However, according to Table~\ref{Tab:C'4vpiF}, the bilinear $\psi^\dagger \1 \psi$ is odd under both time-reversal and charge conjugation, and  $\psi^\dagger \tau^z \psi$  is odd under time reversal. As different components of $u_{ij}$ are invariant under time-reversal and charge conjugation, we conclude that at the leading order  no couplings of spinons to phonons are allowed by symmetry, ${\Hploco}^\text{\piFs}(\bq)=0$.

\begin{table*}
\begin{center}
\begin{tabular}{l c c c c c c ccc cccccc}
\hline\hline
Representation         & $A_2$        & $B_1$       &          $\Bn'_1$ &              $\An'_2$ &            $\En'_1$           &             $\En_3$          &                               $\En_4$          &                                 $\En_5$         \\
Basis& $\tau^z$     &$\tau^z\mu^z$&    $\mu^z$     &                $\1$&$\tau^{x,y},\mu^z\tau^{x,y}$& $\mu^{x,y},\tau^z\mu^{x,y}$&$\mu^{x,y}(\tau^x\mp\tau^y)$&$\mu^{x,y}(\tau^x\pm\tau^y)$ \\
$\cal T$-inv     &  $-$         &    $-$      &      $-$       &              $-$   &             $+$            &              $+$           &                                $-$           &                                 $-$  \\
\hline\hline
\end{tabular}
\caption{\label{Tab:C'4vsF} Explicit form of basis for different irreducible representations of $C_{4v}'$ contained within $G^\text{\sFs}_{\psi^\dagger\psi}$. Action of the group generators coincides with Ref.~\onlinecite{Hermele-sF}. Last row summarizes the transformation of basis elements under time-reversal symmetry. }
\end{center}
\end{table*}

To find a non-zero coupling of spinons to phonons, we allow for the presence of spinon momentum, $\bk$ in the coupling Hamiltonian. This corresponds to the next order in the expansion around the Dirac points. The spinon momentum, $\bk$, transforms under the usual vector representation $E_1$~[this can be inferred from the fact that Dirac Hamiltonian, Eq.~(\ref{Eq:HDiracPiF}) is invariant] and is invariant under time reversal, but odd under charge conjugation. The product  $E_1\times G^\text{$\pi$F}_{\psi^\dagger \psi}$ is reduced~as  
\begin{multline} \label{Eq:E_1^3}
  E_1\times G^\text{$\pi$F}_{\psi^\dagger \psi}
  =
A_1\oplus A_2\oplus  B_1\oplus B_2\oplus  \An'_1 \oplus  \An'_2\oplus \Bn'_1\oplus  \Bn'_2 \\
\oplus 2 (E_1  \oplus \En'_1 \oplus E_2 \oplus \En_3 \oplus \En_4 \oplus \En_5) .
\end{multline}
The first four irreducible representations, which are of interest for us~[cf.\ with Eq.~(\ref{Eq:E1E1C4v})], originate from the  $E_1$ component within $G^\text{$\pi$F}_{\psi^\dagger \psi}$.  Consequently, their basis is analogous to Eq.~(\ref{Eq:E1E1bases}). The  first four irreducible representations  $A_1\ldots B_2$ from Eq.~(\ref{Eq:E_1^3}) can be coupled to corresponding irreducible representations in Eq.~(\ref{Eq:E1E1C4v}). For example, coupling representation $A_1$ with basis $k_x \tau^x+ k_y \tau^y$ to $A_1$ component with basis $u_{xx}+u_{yy}$ results in contribution 
\begin{equation} \label{Eq:Hsp-phPiFExample}
\Hploct^\text{\piFs}
 =
 g^{(1)}_{A_1} (u_{xx}+u_{yy})\big[ k_x \tau^x  +   k_y  \tau^y \big],
\end{equation}
with a phenomenological coupling constant $g^{(1)}_{A_1}$. Dependence on $\bk,\bq$ will be suppressed for brevity in what follows, $\Hploct \equiv \Hploct(\bk,\bq)$. Collecting all contributions at this order, and rearranging phenomenological coupling constants (e.g.\ $g^{(1)}_{1,4} = g^{(1)}_{A_1}\pm g^{(1)}_{B_1}$) we get the most general form of the spinon-phonon interaction Hamiltonian in the $\pi$F phase to be
\begin{multline} \label{Eq:Hsp-phPiF}
   \Hploc^\text{\piFs}
  =
  g^{(1)}_1 (u_{xx}  k_x\tau^x +  u_{yy} k_y \tau^y)
  \\ 
  + g^{(1)}_2 ( u_{xy} k_x\tau^x+ u_{yx} k_y \tau^y)
  + g^{(1)}_3 ( u_{yx} k_x\tau^x +  u_{xy} k_y \tau^y)
  \\ 
  + g^{(1)}_4(u_{xx} k_y \tau^y +  u_{yy} k_x \tau^x).
\end{multline}

\subsection{Results  for the \sFs, \piFk, and  \zFh\ phases \label{SS:PhononsResults}}

After detailed derivation of spinon-phonon interaction for the $\pi$-flux phase on a square lattice, we present results for other algebraic spin liquid phases considered in this work.

\emph{The staggered-flux phase on a square lattice} is similar to the \piFs\ phase, considered above. However, in contrary to the $\pi$-flux phase, there is no charge conjugation present among additional symmetries. As we shall shortly demonstrate, due to reduced symmetry, the number of allowed couplings is going to be larger. Splitting of $G^\text{\sFs}_{\psi^\dagger\psi}$ into irreducible components works as 
\begin{multline} \label{Eq:Gdecomp-sF}
   G^\text{\sFs}_{\psi^\dagger\psi}
  =
  A_2 \oplus B_1 \oplus \Bn'_1 \oplus \An'_2 \oplus 2 \En'_1 \oplus 2\En_3 \oplus \En_4 \oplus \En_5.
\end{multline}
The basis of corresponding components in terms of tensor product of Pauli matrices are listed in Table~\ref{Tab:C'4vsF}. One can notice, that all one-dimensional representation present in the decomposition~(\ref{Eq:Gdecomp-sF}) are odd under time-reversal. Just like the case of the \piFs\ phase, no couplings with phonons are allowed by symmetry at this order. To find non-zero coupling we consider next order in $\bk$. 

\begin{table}[b]
\begin{tabular}{l c c c c c c}
\hline\hline
Representation          & $A_1$        & $A_2$       &          $E_1$ &              $F_1$ &            $F_2$           &             $F_3\&F_4$         \\
Basis& $\1$         &     $\tau^z$&    $\tau^{x,y}$& $\tau^z\mu^{x,y,z}$&               $\mu^{x,y,z}$&        $\tau^{x,y}\mu^{x,y,z}$\\
$\cal T$-inv     &  $+$         &    $-$      &      $-$       &              $+$   &             $-$            &              $+$              \\
\hline\hline
\end{tabular}
\caption{\label{Tab:C6vsFk} Explicit form of basis for irreducible representations of $C'_{6v}$ contained within $G^\text{\piFk}_{\psi^\dagger\psi}$. Notations and action of group generators coincide with those used by Hermele~\emph{et.~al.}~\cite{Hermele-k}}
\end{table}

\begin{table*}
\begin{tabular}{l c c c c c c c c c c c c } \hline\hline
Representation & $A_1$ & $B_1$ & $A_2$ & $B_2$ & $E_1$ & $E_2$ &
$A_1$ & $B_1$ & $A_2$ & $B_2$ & $E_1$ & $E_2$ \\ 
Basis & $\1$ & $\mu^z$ &
$\tau^z$ & $\mu^z\tau^z$ & $\tau^{x,y}$ &
$-\mu^z\tau^y,\mu^z\tau^x$ &
$\mu^x\tau^z$ & $\mu^y\tau^z$ & $\mu^x$ & $\mu^y$
& $\mu^x\tau^y,-\mu^x\tau^x$ &
$\mu^y\tau^{x,y}$ \\ 
$\calT$-inv & $+$ & $-$ & $-$ & $+$ & $-$ & $+$ & $+$ & $+$ &
$-$ & $-$ & $+$ & $+$ \\ \hline\hline
\end{tabular}
\caption{\label{Tab:GC6v}  Irreducible representations of $C_{6v}$ contained within $G^\text{\zFh}_{\psi^\dagger\psi}$ and their basis. Each irreducible component occurs twice: first six representations in the Table are diagonal in the valley space,  whereas remaining six are their off-diagonal counterparts.  Adopted from Table~III in Ref.~\onlinecite{Basko}.
}
\end{table*}

Explicit expression for $\Hploct(\bk)$ can be found using decomposition of the  product $\En'_1\times  G^\text{\sFs}_{\psi^\dagger\psi}$~[where $\En'_1$ corresponds to spinon momentum~\footnote{%
Naively, the fact that $\bk$ transforms under $E_1'$ rather than $E_1$ seems to contradict the expectation, that as the staggered flux becomes exactly equal to $\pi$, the \sFs\ phase continuously evolves into the \piFs\ phase. In reality, there is no contradiction: the \sFs\ phase at the value of flux of $\pi$ indeed turns into the \piFs\ phase, however, realized with a different ansatz.%
}] into irreducible representations,
\begin{multline} \label{Eq:Gdecomp-sFE3}
\En'_1\times  G^\text{\sFs}_{\psi^\dagger\psi}
=
2(A_1\oplus A_2 \oplus  B_1 \oplus B_2 \\
\oplus E_1 \oplus \En'_1 \oplus E_2 \oplus  \En_3 \oplus \En_4 \oplus \En_5).
\end{multline}
The overall factor of two indicates that there are two distinct copies of each irreducible representation in the decomposition. Four one dimensional irreducible representation in the first line of Eq.~(\ref{Eq:Gdecomp-sFE3}) coincide with the decomposition of $E_1\times E_1$ into irreducible representations. As all irreducible components are encountered twice in the decomposition~(\ref{Eq:Gdecomp-sFE3}), we have eight different couplings between spinons and phonons.

The basis for the two copies  one-dimensional representations $A_1\ldots B_2$ in Eq.~(\ref{Eq:Gdecomp-sFE3})  can be deduced using the fact that all these irreducible representations originate from the tensor product $\En'_1\times \En'_1$:
\begin{subequations} \label{Eq:E3E3basis}
\begin{eqnarray} 
  A_1 : k_x \tau^x+k_y \tau^y, 
 \qquad
 A_2 : k_x \tau^y-k_y \tau^x ,
 \\
 B_1 : k_x \tau^x-k_y \tau^y,
 \qquad
 B_2 :  k_x \tau^y+k_y \tau^x.
\end{eqnarray}
\end{subequations}
The bases for the second copy of irreducible representations in Eq.~(\ref{Eq:Gdecomp-sFE3}) have the same form as in Eq.~(\ref{Eq:E3E3basis}), but with an extra Pauli matrix $\mu^z$. Physically, this corresponds to the fact that the anisotropy in Fermi velocity is not prohibited by symmetry in the \sFs\ phase.~\cite{Hermele-sF} Time-reversal invariance does not reduce the number of allowed couplings, as all combinations in Eq.~(\ref{Eq:E3E3basis}) are invariant under $\calT$. The resulting spinon-phonon interaction Hamiltonian, has eight terms, four of which coincide with the case of the \piFs\ phase, Eq.~(\ref{Eq:Hsp-phPiF}), while remaining four terms contain an extra $\mu^z$ and correspond to a phonon-induced valley anisotropy in a Fermi velocity.

\emph{The $\pi$-flux phase on a kagome lattice} has a symmetry group $C'_{6v}$. Just like in the case of square lattice, it is an extension of a conventional symmetry group of hexagon, $C_{6v}$, to the case of enlarged unit cell.~\cite{Hermele-k} The point group of hexagon governs the properties of phonons. The  generators of $C_{6v}$ are rotations for an angle of $\pi/3$, $\Rot_{\pi/3}$, and reflection with respect to $x$-axis,~$\Ref_x$. Using character Table~\ref{Tab:C6v}, the tensor product of two vector representations is decomposed as
\begin{equation}\label{Eq:C6vE1E1}
  E_1\times E_1
  =
  A_1 \oplus A_2 \oplus E_2.
\end{equation}

On the other hand, using character table for the group $C'_{6v}$ from Ref.~\onlinecite{Hermele-k}, one can reduce representation on fermion bilinears as
\begin{equation}\label{Eq:C6vG}
  G^\text{\piFk}_{\psi^\dagger\psi}
  =
  A_1 \oplus A_2 \oplus E_1 \oplus F_1 \oplus F_2 \oplus F_3 \oplus F_4.
\end{equation}
Comparing Eqs.~(\ref{Eq:C6vE1E1}) and (\ref{Eq:C6vG}), we see that, in principle,  couplings between corresponding $A_1$ and $A_2$ components are possible.  However, if we consider properties of representations under time reversal, listed in Table~\ref{Tab:C6vsFk}, we see that basis of $A_2$ is odd under time reversal and only  coupling to density fluctuations  remains:
\begin{equation} \label{Eq:piFk-density}
  \Hploco^\text{\piFk}
  =
  g^{(0)}_0 (u_{xx}+u_{yy})\1.
\end{equation}
This coupling vanishes for transverse phonon modes, and we consider terms which are next order in $\bk$.  Higher order terms can be found from decomposing of  $E_1\times G_{\psi^\dagger\psi}^\text{\piFk} $ into irreducible components,
\begin{equation}\label{Eq:E1GpiFk}
  E_1\times G_{\psi^\dagger\psi}^\text{\piFk}
  =
  A_1 \oplus A_2 \oplus 2E_1 \oplus E_2 \oplus 2(F_1 \oplus F_2 \oplus F_3 \oplus F_4),
\end{equation}
and pairing those with irreducible representations contained within $E_1\times E_1$, Eq.~(\ref{Eq:C6vE1E1}).
Only irreducible representations $A_1$, $A_2$ and $E_2$ in Eq.~(\ref{Eq:E1GpiFk}) are of interest as the same components are also present in decomposition~(\ref{Eq:C6vE1E1}). The basis for these representations can be written in terms of basis of $E_1$, $(k_x,k_y)$ and $E_1$ component within $G_{\psi^\dagger\psi}^\text{\piFk}$, [$\btau=(\tau^x,\tau^y)$, see Table~\ref{Tab:C6vsFk}]:
\begin{subequations}
\begin{eqnarray}
  A_1
  &: &
  \bk\cdot\bm \tau 
  \\
  A_2
 & : &
  \bk\times\bm\tau
  \\
 E_2
  &: &
  (k_x\tau^y+k_y\tau^x,k_x\tau^x-k_y\tau^y).
\end{eqnarray}
\end{subequations}
From here, we can read off the most general form of the spinon-phonon interaction Hamiltonian to be 
\begin{multline} \label{Eq:Hsp-phPiFk}
   \Hploc^\text{\piFk}
  =
  g^{(1)}_1 (u_{xx}+u_{yy})(\bk\cdot\bm\tau)
  + g^{(1)}_2 (u_{xy}-u_{yx})( \bk\times\bm\tau)
  \\
  + g^{(1)}_3  \big[ (u_{xy}+u_{yx})(k_x\tau^y+k_y\tau^x)+  (u_{xx}-u_{yy})(k_x\tau^x-k_y\tau^y)\big].
\end{multline}

\emph{The uniform phase on a honeycomb lattice} is also governed by the symmetry group $C_{6v}$. The notable difference compared to the cases considered above, is that the honeycomb lattice is not Bravais and has a unit cell consisting of two atoms. Derivation of spinon-phonon interaction is analogous to the case of graphene.~\cite{Basko} We neglect the optical phonon modes related to the presence of two atoms in the unit cell.  Tensor product of two vector representations is given by Eq.~(\ref{Eq:C6vE1E1}). Whereas, decomposition of $G_{\psi^\dagger\psi}^\text{\zFh}$ into irreducible representations works as:~\cite{Basko}
\begin{equation}\label{Eq:G0Fh}
 G_{\psi^\dagger\psi}^\text{\zFh}
  =
  2 (A_1 \oplus A_2 \oplus B_1 \oplus B_2 \oplus E_1 \oplus  E_2),
\end{equation}
where two copies describe matrices diagonal and non-diagonal in valley space.  Bases of different irreducible components are given in Table~\ref{Tab:GC6v}. Terms which are off-diagonal in the valley space are not considered, as we do not allow for the intervalley scattering due to phonons.  At the zeroth order in $\bk$, in addition to the density coupling, Eq.~(\ref{Eq:piFk-density}),  there are terms which do not vanish for transverse phonon modes,
\begin{multline} \label{Eq:Hsp-ph0Fh}
   \Hploc^\text{\zFh}
  =
  g_1 \big[(u_{xx}- u_{yy})\tau^x-( u_{xy}+ u_{yx})\tau^y\big]\mu^z.
\end{multline}
Thus there is no need to consider next order in $\bk$. Note, that as basis of $A_2$ is odd under time reversal, we omit otherwise possible term $(\p_x u_y - \p_y u_x) \tau_z$ from Eq.~(\ref{Eq:Hsp-ph0Fh}).

\subsection{Comparison between different phases}

It is instructive to compare the above results for the spinon-phonon interaction in different realizations of Dirac spin liquid. In all derivations we considered Dirac fermions describing low energy excitations. The Dirac dispersion arises as an approximation of the band structure in vicinity of $K_\pm$ points in the Brillouin zone. Consequently, the interaction with acoustic phonons may be understood from the influence of lattice deformations on the low energy band structure.  The coupling of phonons to the density of spinons is very easy to explain from this perspective. The local changes in the volume of the lattice, described exactly by $\mathop{\rm div} \bu = u_{xx}+u_{yy} $, correspond to the density modulations of spinons, yielding the interaction Hamiltonian~(\ref{Eq:piFk-density}). In the case of the  \zFh\ phase, remaining terms given by Eq.~(\ref{Eq:Hsp-ph0Fh}) are can be interpreted as a relative shift of $K_\pm$ points with respect to each other by lattice deformations. In other words, strain is translated into a gauge field, which coupled with opposite sign in different valleys -- well known effect for the case of graphene.~\cite{Basko,Vozmediano}

The presence of fluxes and non-trivial action of projective symmetry group prohibits density coupling for \piFs\ and \sFs\ phases. In the \piFk\ phase, the density coupling is the only allowed coupling at this order. To find a non-trivial couplings, we considered next order expansion in vicinity of the Dirac points. These couplings may be readily understood as a \emph{deformation} of the band structure in vicinity of $K_\pm$ point, which, nevertheless leaves the position of the Dirac point within the Brillouin zone intact. This is exactly what we see in couplings~(\ref{Eq:Hsp-phPiF}) and (\ref{Eq:Hsp-phPiFk}), which can be interpreted as the change in Fermi velocity,~$\vf$.  Note, that the position of Dirac points in the Brillouin zone is non-universal, and depends on the choice of the implementation of the given phase.  Therefore, it is natural, that the (physical) lattice deformation does not have any impact on the (unphysical) position of the Dirac points.

\section{Sound attenuation \label{S:attn}}

We continue with a discussion of observable consequences of spinon-phonon interaction. Interaction of acoustic phonons with gapless spinons opens another channel for decay of phonons. Thus, it is expected to contribute to the attenuation of ultrasound. To get an estimate of the this effect, we perform a simple calculation in this Section. As an example, we consider the algebraic spin liquid phase with staggered flux on a square lattice. We comment on the differences  for the \piFk\ and \zFh\ phases.  We do not consider the \piFs\ phase, to avoid the complications related to the presence of an SU(2) gauge field.

\subsection{Framework}
\begin{figure}
\begin{center}
\includegraphics[width=0.85\columnwidth]{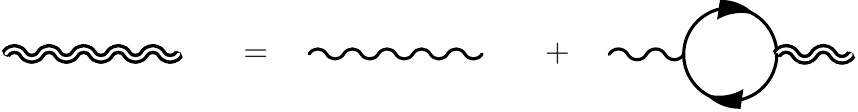}\\
\caption{ \label{Fig:GaugeProp} Double wavy line shows the gauge field propagator in the RPA approximation. Thin wavy line is the bare Maxwell propagator.}
\end{center}
\end{figure}

We start with establishing the framework and introducing the basic elements required to calculate the ultrasound attenuation. These are the gauge field propagator and the phonon self-energy.

The gauge field, strongly coupled to spinon emerges from a microscopic constraint and fluctuations around mean field anzatz. On a microscopic level it originates from constraint and is not dynamical. Non-trivial dynamics of the gauge field is generated due to the coupling to fermions.~\cite{IoffeLarkin,KimLee} First, the Maxwell term will be generated  while integrating out high energy degrees of freedom. Another contribution, which is singular compared to the Maxwell term, comes from the Dirac band touching and can be written via vacuum polarization operator for massless Dirac fermions. Total action of gauge field then becomes
\begin{equation} \label{Eq:SGaugePi}
  S_{a}
  =
  \frac12 \int \frac{d^3\vq }{(2\pi)^3}a_\mu(\vq) [(D^\text{M})^{-1}_{\mu\nu}(\vq)-\Pi_{\mu\nu}(\vq)]a_\nu(-\vq),
\end{equation}
where we use covariant notations in Euclidean space, $\vq = (i\omega,\vf\bq)$.   $\Pi_{\mu\nu}(\vq)$ is the polarization bubble of Dirac fermions (see Fig.~\ref{Fig:GaugeProp}), and $(D^\text{M})^{-1}_{\mu\nu}$ is the inverse Maxwell propagator of the gauge field~(we work in the Lorentz gauge, $\vk\cdot \vec a =0$), 
\begin{equation} \label{Eq:DMinvmunu}
 (D^\text{M})^{-1}_{\mu\nu}(\vq)
  =
  \left(\delta_{\mu\nu} - \frac{\vq_\mu \vq_\nu}{\vq^2}\right) \Pi^\text{M}(\vq),
\end{equation}
where  $\Pi^\text{M}(\vq)$ corresponds to the inverse propagator without tensor structure:
\begin{equation} \label{Eq:DMinv}
   \Pi^\text{M}(\vq) =
   \frac{\vq^2}{e^2}.
\end{equation}
The polarization bubble at zero temperature~(projector tensor structure is again omitted) is given by:~\cite{Dorey,KhveschenkoPRB06,VafekThesis} 
\begin{equation} \label{Eq:Polarization}
  \Pi(\vq)
  =
  -\frac{N}{8} \sqrt{\vq^2}
  ,
\end{equation}
where we introduced the integer number of flavors of four-component Dirac fermions, $N$, in our theory. The physical case corresponds to $N=2$ coming from spin. 
Action~(\ref{Eq:SGaugePi}) translates into the total propagator for the gauge field, Fig.~\ref{Fig:GaugeProp}, given by
\begin{equation} \label{Eq:GaugeProp}
  D(q)
  =
  \frac{8}{N}\frac{1}{\sqrt{\vq^2}+{8\vq^2}/({Ne^2})}.
\end{equation}
In what follows we will need the polarization bubble at finite temperature, which may be written as
\begin{equation} \label{Eq:Pi-fin-T}
  \Pi_{\mu\nu} 
  =
  A_{\mu\nu}\Pi^A+   B_{\mu\nu}\Pi^B.
\end{equation}
The tensors $A_{\mu\nu}$ and $B_{\mu\nu}$,
\begin{eqnarray}
  \label{Adef}
  A_{\mu\nu}
  &=&
  \left(\delta_{\mu0}-\frac{q_\mu q_0}{\vq^2}\right)
  {\displaystyle\frac{\vq^2}{\bq^2}}
  \left(\delta_{0\nu}-\frac{q_0 q_\nu}{\vq^2}\right),
  \\ \label{Bdef}
  B_{\mu\nu}
  &=&
  \delta_{\mu i}\left(\delta_{\mu0}
  -
  \frac{q_\mu q_0}{\vq^2}\right)\delta_{j\nu},
\end{eqnarray}
are orthogonal to each other and their sum reproduces the original zero-temperature tensor structure, Eq.~(\ref{Eq:DMinvmunu}).
Explicit expressions for $\Pi^{A,B}$ along with detailed calculations are available in the literature.~\cite{VafekThesis} We need the asymptotic expression for $\Pi^{A}$  in the limit $T\gg v_F|\bq|\gg \om $:
\begin{equation} \label{Eq:PiA-T}
  \Pi^A
  =
  -\frac{2N T\log 2}{\pi}\left(1+\I\frac{\om}{\vf q}\right).
\end{equation}

Sound attenuation, $\alpha_\text{s}$, will be calculated from the self-energy of phonons, $\Pi_\text{ph}(\om,\bq)$, arising due to interactions with spinons. More precisely, $\alpha_\text{s}$ is given by the imaginary part of the retarded self-energy,~\cite{ZhouLee}
\begin{equation} \label{Eq:alphas}
\alpha_\text{s}
=
-\frac{2}{v_s} \Im \left[\Pi^R_\text{ph}(\om,|\bq|)\right]_{\om=v_s|\bq|},
\end{equation}
with frequency and momentum related by the dispersion relation of the acoustic phonons, $\om=\vs|\bq|$, where $\vs$ is the sound velocity. 

Let us discuss the approximations to be used in the calculation of the sound attenuation. For simplicity, we consider the clean case, i.e.\ we assume that the mean free path of spinons, $l$, is much larger than the ultrasound wavelength, $q l \gg 1$. Also, we assume that the sound velocity is much smaller than the Fermi velocity, $\vf\gg \vs$. Under this condition, we immediately find that if $\om$ and $q$ are the phonon energy and the wave vector,  $\vf q  = (\vf/\vs) \om \gg \om$. Finally, in contrary to the case of spin liquid with a Fermi surface,~\cite{ZhouLee} non-zero temperature is required to get non-vanishing sound attenuation in a Dirac spin liquid.  This is a consequence of the energy and momentum conservation in the scattering process. Acoustic phonon cannot excite a particle-hole pair of spinons since the maximum momentum change for such a pair with energy $\om$, $\Delta k  = \om/\vf$, is much smaller than phonon momentum, $q = \om/\vs$. Therefore, we assume that the system is at a finite temperature $T\gg (\vf/\vs)\om\gg \om$.

As noted above, there is a gauge field strongly coupled to the spinons. In order to have a control over its effects we artificially introduced the number of flavors, $N$, being equal to two in the physical case. Since gauge field propagator, Eq.~(\ref{Eq:GaugeProp}), is proportional to $1/N$, effects of gauge field are suppressed for large $N$. We will perform calculations of sound attenuation to the leading order within $1/N$ expansion, commenting on the higher order terms.

\subsection{Sound attenuation in  the \sFs\ phase}
\begin{figure}
\begin{center}
\includegraphics[width=0.99\columnwidth]{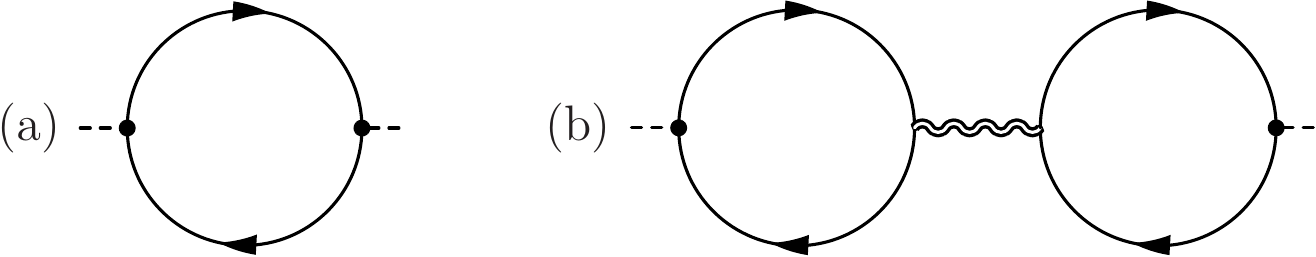}\\
\caption{ \label{Fig:Pi-L} Contribution of spinons to the longitudinal sound attenuation. The bare contribution from spinons is given by the diagram (a).  Diagram (b) accounts for the screening  due to fluctuations of the  gauge field. Black dots represent spinon-phonon interaction vertex, specified in the main text.}
\end{center}
\end{figure}

Having the basic ingredient for the calculation of sound attenuation at our disposal, we consider $\alpha_s$ for longitudinal phonons in the \sFs\ phase. As shown in Section~\ref{SS:PhononsPiF}, there is no allowed coupling at the leading order in $\bk$. All possible couplings at the next order are given by Eq.~(\ref{Eq:Hsp-phPiF}). For simplicity, we consider only first term in Eq.~(\ref{Eq:Hsp-phPiF})~[see Eq.~(\ref{Eq:Hsp-phPiFExample})]. Combining Eqs.~(\ref{Eq:u(r)}) and (\ref{Eq:Hsp-phPiFExample}), corresponding spinon-phonon interaction vertex reads:
\begin{subequations}\label{Mkxkx}
\begin{eqnarray} \label{Mkxkx1}
  M^{(1)}_{\bk}(\bq)
  &=&
    {\tg}^{(1)}_1  q\,   \tilde  M^{(1)}_{\bk}(\hat \bq),
\\ \label{Eq:tildeMkxkx}
 \tilde  M^{(1)}_{\bk}(\hat \bq)
 &=&
 \hq_x^2 k_x\tau^x+\hq_y^2 k_y \tau^y,
\end{eqnarray}
\end{subequations}
where $\hat \bq = (\hq_x,\hq_y)$ is the unit vector pointing along $\bq$. In what follows, coupling constants with tilde are defined~as:
\begin{equation} \label{Eq:tg-def}
  \tg 
  = 
  \sqrt{\frac{1}{2\rho \om_\bq}} g.
\end{equation}

To leading order in $1/N$, the polarization operator of phonons due to interaction with spinons is given by the sum of two diagrams in Fig.~\ref{Fig:Pi-L} with the spinon-phonon interaction vertex from Eq.~(\ref{Mkxkx}). Indeed, the first diagram in Fig.~\ref{Fig:Pi-L} has one fermionic bubble and is proportional to $N$. An extra fermionic bubble in the diagram Fig.~\ref{Fig:Pi-L}~(b) is compensated by factor of $1/N$ from the gauge field propagator. We have for the first contribution, Fig.~\ref{Fig:Pi-L}~(a):
\begin{equation} \label{Eq:as-1}
  \alpha_s
  =
  \frac{2}{v_s} \big[{\tg}^{(1)}_1\big]^2 q^2 \Im \Pi^{R(1)}(\om,\bq).
\end{equation}
The imaginary part of the bubble diagram with spinons is calculated in the Appendix~\ref{App:ImPi} and behaves as $\Im \Pi^{R(1)}(\om,\bq)\propto N \omega  T^3/(\vf^3 q)$ at the leading order. One can show that the contribution from the diagrams with an extra gauge field propagator, Fig.~\ref{Fig:Pi-L}~(b), has the same order of magnitude as Fig.~\ref{Fig:Pi-L}~(a). Thus, using  dispersion relation of acoustic phonons, we get the following estimate for the sound attenuation:
\begin{equation} \label{Eq:as-2}
  \alpha_s
  \propto
  N \big[{\tg}^{(1)}_1\big]^2
  \frac{\om^4 T }{\vs^2 \vf^3}. 
\end{equation}
The value of the coupling constant, $g^{(1)}_1$, may be estimated from the sensitivity of the velocity of Dirac spinons to the changes of the lattice constant, $a$:
\begin{equation} \label{Eq:g-est}
  g^{(1)}_1 \sim a \frac{\p \vf}{\p a} 
  \sim \vf.
\end{equation}
Using this estimate, we obtain for the sound attenuation:
\begin{equation} \label{Eq:as-est}
  \alpha_s
  \sim N \left(\frac{T}{\om_D}\right)^2 \alpha^{(0)}_s,
\end{equation}
where $\alpha^{(0)}_s$ defined as
\begin{equation} \label{Eq:asFL-est}
  \alpha^{(0)}_s
  \sim q \frac{k_T}{m_\text{ion} \vs} = \frac{qT}{m_\text{ion} \vs\vf}.
\end{equation}
The Debye frequency has been estimated as $\om_D \propto \vs/a$, and $k_T=T/\vf$ is a wave vector of spinons with the energy equal to the temperature. The $\alpha^{(0)}_s$, introduced above,   gives the estimate for the sound attenuation coefficient in the case of spinon Fermi surface if one substitute the Fermi momentum for the $k_T$, $k_T \to k_F$.  We see that in a Dirac spin liquid, contribution of spinons to the sound attenuation is suppressed compared to the Fermi surface case by two factors. The first factor,  $T/\mu_F \ll 1$ is generic for any Dirac spin liquid and originates from the  vanishing density of states at zero temperature in the Dirac spectrum. The second factor $(T/\om_D)^2$, which is also expected to be smaller than one, arises due to peculiar form of spinon-phonon coupling. 

Finally, we comment on the next order in $1/N$ terms, contributing to the sound attenuation.  There is a much larger number of diagrams at the order $O(1)$. The most obvious are the vertex corrections, where gauge field dresses the interaction vertex of spinons with phonons or gauge field itself. Note, that if the spinon-phonon interaction vertex corresponded to some conserved current, it would be protected from logarithmic corrections.~\cite{KimLee} However, this seems to be not the case here and in general we expect logarithmic corrections to arise at the order $O(1)$. There is also another type of contribution $O(1)$, which is more unusual.  Indeed, in order to maintain the gauge invariance, $\bk$ in Eq.~(\ref{Mkxkx}) has to be extended to include the gauge field as well. This leads to the vertex where a phonon can generate a quanta of the gauge field \emph{in addition} to the particle-hole pair of spinons. Similar type of vertex has been considered in Ref.~\onlinecite{KimPRB94}. 

\subsection{Sound attenuation in \piFk\  and \zFh\ phases}

As we have shown above, the peculiar form of the coupling between phonons and spinons in the \sFs\ phase leads to the suppression of the sound attenuation coefficient by additional small factors. 
One may expect, that since in the  \piFk\  and \zFh\ phases longitudinal phonons couple to the density of spinons, the sound attenuation will be parametrically larger than  for the \sFs\ phase. Below, we are going to demonstrate that these naive expectations do not hold. Due to the presence of the gauge field, the density coupling gets screened and does not contribute to the sound attenuation at leading order in~$1/N$. 

Using explicit form of the coupling, Eq.~(\ref{Eq:piFk-density}), we write corresponding spinon-phonon interaction as:
\begin{equation} \label{Eq:M-dens}
  M^{(0)}_{\bk}(\bq)
  =
  \tg^{(0)}_0  q\, \1.
\end{equation}
The identity matrix corresponds to the spinon density, thus justifying the use of the term ``density couping''. The self-energy of phonons, required to find the sound attenuation, can be expressed via time component of electron polarization bubble $\Pi_{00}$. In addition, one have to account for the effect of gauge field, including the scalar potential~(unlike the case of spinons with Fermi surface, scalar potential of the gauge field is not screened by Dirac fermions). These two contributions to $\Pi_\text{ph}$ are shown in Fig.~\ref{Fig:Pi-L}, where now black dots correspond to the interaction vertex~(\ref{Eq:M-dens}). Accounting for the both diagrams in Fig.~\ref{Fig:Pi-L}, we get:
\begin{equation} \label{Eq:Pi-L1}
  \Pi_\text{ph} (\vq)
  =
  \big[\tg^{(0)}_0\big]^2 q^2
  \big[\Pi_{00}(\vq)+\Pi_{0\mu}(\vq) D_{\mu\nu}(\vq) \Pi_{\nu 0}(\vq) \big],
\end{equation}
where propagator and self-energy are taken on a phonon mass shell, $\omega = v_s |\bq|$. Using the finite-temperature expression for the polarization operator, Eq.~(\ref{Eq:Pi-fin-T}), we find that only $\Pi^A$ contributes in the present case. Two terms in the sum in Eq.~(\ref{Eq:Pi-L1}) partially cancel each other and we arrive at:
\begin{equation} \label{Eq:Pi-L2}
  \Pi_\text{ph}(\vq)
  =
    \tg_0^2 q^2 \frac{q^2}{\vq^2} 
  \frac{\Pi^\text{M}(\vq)\Pi^A(\vq)}{\Pi^\text{M}(\vq) - \Pi^A(\vq)}.
\end{equation}
Since $\Pi^A(\vq)$ is more important than the Maxwell term, at the leading order we can neglect the latter term in the denominator, and get $\Pi_\text{ph}(\vq)\propto -\Pi^\text{M}(q)$. Note, that this term is of order of $O(1)$, compared to the naive expectation $\Pi_\text{ph}(\vq) \sim O(N)$. Moreover, this term does not contribute to the imaginary part of the self-energy: Maxwell propagator originates from high-energy modes, whereas we are interested in the decay of phonons into low-energy Dirac-like spinons. Omitting the leading order term, and including next order contribution, we get the result
\begin{equation} \label{Eq:Pi-L3}
   \Pi_\text{ph}(\vq)
  =
   -     \tg_0^2 q^2 \frac{q^2}{\vq^2}\frac{[\Pi^\text{M}(\vq)]^2}{\Pi^A(\vq)},
\end{equation}
which is proportional to $1/N$. Qualitatively, cancellation of two leading terms can be understood as an effect of screening due to gauge field. 

Now that we have shown that the contribution of the density coupling to the sound attenuation is proportional to $1/N$ and thus negligible, we consider other terms in the coupling Hamiltonian, contributing at the order $O(N)$. For the \piFk\ phase, these terms, listed in Eq.~(\ref{Eq:Hsp-phPiFk}), are first order in $\bk$. Thus, the sound attenuation is expected to be of the same order as the results for the \sFs\ phase, listed in Eq.~(\ref{Eq:as-est}). 

However, for the \zFh\ phase there are couplings allowed without an extra $\bk$, see Eq.~(\ref{Eq:Hsp-ph0Fh}). Contribution from these couplings is expected to be of order of $\alpha^{(0)}_s$~[see Eq.~(\ref{Eq:asFL-est})]. We note that contribution of gauge field vanishes in the present case. Indeed, it couples with an opposite sign in different valleys [note the presence of the  extra $\mu^z$ matrix in Eq.~(\ref{Eq:Hsp-ph0Fh})], thus diagram in Fig.~\ref{Fig:Pi-L}~(b) is identically zero.

\section{Discussion and outlook\label{S:discuss}}

We presented the general procedure for the derivation of the coupling between spinons and acoustic phonons in the Dirac spin liquid. Our procedure is based on the symmetry arguments. Although general fermionic bilinear transforms under projective representation of the lattice symmetry group, spin singlet bilinears realize conventional (i.e., not projective) representation of the microscopic symmetry group.  We found the decomposition of this representations into irreducible for  $\pi$-flux and staggered flux phases on a square lattice, as well as for $\pi$-flux phase on kagome lattice and a Dirac spin liquid phase on a honeycomb lattice. By pairing corresponding irreducible representations with those for acoustic phonons, we were able to identify all symmetry allowed couplings. Note, that such decomposition can have other applications. For instance, it can be used to derive symmetry allowed couplings to optical phonons or some other excitations.

In a continuum limit all considered spin liquid phases have similar low energy Dirac excitations, and hardly can be distinguished. Nevertheless, the  allowed interactions with phonons have different form. For the Dirac spin liquid phase on a honeycomb lattice the coupling to acoustic phonons is similar to the case of graphene. The only difference is that the coupling to the density of spinons, naively expected to be the largest, is screened by the gauge field~(this is true for all Dirac spin liquid phases). As a result, for  spin liquid phases on a square and kagome lattices  considered in this work, the leading couplings contains an extra small parameter $(T/\om_D)^2$, compared to U(1) Dirac spin liquid on honeycomb lattice. Qualitatively, in these phases, the lattice deformations with small wave vectors couple to the changes or anisotropies in Fermi velocity. Whereas in the case of zero-flux phase on honeycomb lattice such lattice deformations shift the position of Dirac points, acting similarly to the gauge~field.

The difference between couplings arises naturally from the fact that they are controlled by the representation of the corresponding symmetry group, acting on a lattice level.  Thus the interaction of spinons with phonons retains some information about microscopic structure of the phase. It would be instructive to check if one can distinguish between different  projective realizations of the same symmetry group by looking at couplings to spinons. The simplest example~\footnote{X.-G.Wen, private communication.} of such two phases are  two $\mathds{Z}_2$ spin liquid phases on a square lattice~(Z2A0013 and Z2Azz13 in notations of Refs.~\onlinecite{Wen-PSG,Wen-PSG2}). This, however, requires generalization of the present approach to the case of $\mathds{Z}_2$ spin liquid phases, which is an interesting open question. Another open question is to understand the effect of projective realization of spin SU(2) symmetry, which has been proposed recently.~\cite{HermelePRB11}

In order to understand the perspectives of spinon-phonon interaction as a probe of fermionic spinons, we carried out a simple calculations within $1/N$ expansion. Assuming that our results can be extrapolated to the physical case $N=2$, we see that the in a generic Dirac spin liquid exemplified by the zero flux phase in the honeycomb lattice, sound attenuation is suppressed due to vanishing density of states at zero temperature and $\alpha_S \propto q T/(m_\text{ion}\vf\vs )$. On the other hand, the peculiar form of spinon-phonon coupling  in the $\pi$-flux and staggered flux phases contributes an additional   suppression of the form~$(T/\om_D)^2$. Nevertheless, the effect from phonons is still potentially observable, as the sound attenuation due to phonon-phonon scattering (caused by non-liearities) behaves as $\alpha\sim T^4$ for  $T\ll \omega_D$.~\cite{Woodruff}

We note that our calculations should be viewed as a simple estimate due to the nature of approximations used. Currently, to the best of our knowledge there is no experimental data available on the sound attenuation in Dirac spin liquids. Provided such data becomes available, more extensive theoretical work is required, in order to construct a realistic description. In particular, for the prospective spin liquid phase on a kagome lattice,\cite{Helton07,Han12} the clean limit, assuming mean free path $l\gg q^{-1}$ does not apply. Also, the estimate for $\vf$ suggests that $\vf \sim \vs$, rather than $\vf\gg \vs$ as was assumed.
Another question, which can be  relevant for a spin liquid on a kagome lattice is the effect of transition from U(1) to $\mathds{Z}_2$ spin liquid and its possible manifestation in the ultrasound attenuation. 

\section*{Acknowledgments}
M. S. is grateful to X.-G. Wen, L. Levitov, M. Metlitski, K. Michaeli, K.-T. Chen, and  A. Potter for many useful discussions. We acknowledge  support by grant NSF DMR 1104498. We acknowledge the hospitality of KITP, where final stages of this project were completed.

\appendix
\section{Elements of representation theory for relevant groups \label{A:repr}}
This Appendix  provides background on the representation theory, and gives more details for the symmetry groups used in the main text.  It starts with a summary of the basic facts from the representation theory of finite groups, which are extensively used throughout the paper. The reader interested in more details or derivations of particular statements is referred to Refs.~\onlinecite{GroupTheory,GroupTheoryOnline}. Next, the symmetry group of square and its extension, relevant for the \piFs\ and \sFs\ phases, is considered. Finally, the basic facts about the symmetry group of hexagon and the symmetry group of the \piFk\ phase are discussed.

\subsection{Basic facts from representation theory}

We consider a point group $\calG$, which contains $h_\calG$ elements. Notion of conjugacy classes will be of great importance  for us in what follows.  Conjugacy class is defined as a complete set of mutually conjugate group elements, where two group elements $g_1$ and $g_2$ are defined to be conjugate if there exists another group element $g_3$, such that $g_1 = g_3^{-1} \circ g_2 \circ g_3$. In other words, if $g$ belongs to a given conjugacy class, $\calC_i$, then for any group element 
\begin{equation} \label{conj-def}
\forall\ g_j \in \calG, \qquad  g_j^{-1} \circ g \circ g_j \in \calC_i,
\end{equation} 
still is an element from the conjugacy class $\calC_i$. Let us assume, that the group $\calG$ has $n_\calG$ conjugacy classes, denoted as $\calC_1, \calC_2,\ldots \calC_{n_\calG}$. Each class contains $N_k$ elements, and, since each group element belongs to only one conjugacy class, we have  $\sum_{k=1}^{n_\calG} N_k = h_\calG$. The identity element, which is necessary present in any group is a conjugacy class itself, $\calC_1 \equiv E =\{\1\}$ and $N_1=1$. For an abelian group, any element belongs to a separate conjugacy class, so that $n_\calG = h_\calG $, and $N_{1,\ldots, n_\calG}=1$.

In what follows, our main interest will be in classifying representations of a given group. Representation of the group can be thought of as a mappings from the group elements to operators acting on some linear space, $g \to R_g$ which respects the group multiplication, $R_{g_1} \cdot R_{g_2} = R_{g_1\circ g_2}$. If operators from a given representations cannot be represented as a direct sum of two operators acting on a smaller subspaces, this representation is called irreducible. According to this definition, any representation $D$ can be expressed as a direct sum of irreducible representations,
\begin{equation} \label{Eq:Gtoirrep}
  D
  =
  a_1 D^{(1)} \oplus  a_2 D^{(2)} \oplus \ldots  \oplus a_{n_\calG} D^{(n_\calG)},
\end{equation}
where non-negative integers $a_i$ describe how many times a given irreducible representation is encountered in the decomposition. If $D^{(i)}$ is not contained within $D$, corresponding $a_i$ is zero, $a_i = 0$.
In this way the problem of classifying all representations of a given group is reduced to a classification of all irreducible representations. 

The number of different irreducible representations for the group $\calG$ coincides with the number of its conjugacy classes, $n_\calG$. Each irreducible representation, $D^{(i)}$ is specified by the value of its character for different conjugacy classes, defined as 
\begin{equation} \label{Eq:chi-def}
\chi^{(i)}(\calC_k) = \tr R_{g_{\calC_k}},
\quad
\text{where}
\quad
g_{\calC_k} \in \calC_k.
\end{equation}
According to the definition of the conjugacy class~(\ref{conj-def}), the value of $\chi^{(i)}(\calC_k)$ does not depend on the choice of a particular element $g_{\calC_k}$  from the $\calC_k$. Operators which act on a linear space can be expressed as matrices, and trace in Eq.~(\ref{Eq:chi-def}) is understood in this sense. 

Value of character for the conjugacy class which consists identity  $E =\{\1\}$ is special, since it gives us the dimension of the corresponding irreducible representation. To classify all irreducible representations of a given group,  it is good to know not only the number of different irreducible representations, but their dimensions as well. In such situation the following relation between the number of elements in the group, $h_\calG$ and the dimensions of all irreducible representations contained within the group, $\ell_i = \chi^{(i)}(E)$ turns out to be particularly useful:
\begin{equation} \label{Eq:dim-irreps}
  h_\calG
  =
  \sum_{i=1}^{N_\calC} \ell_i^2.
\end{equation}
Typically only a few sets of integers $\{\ell_1,\ldots,\ell_{N_\calG}\}$ satisfy this relation, and one can usually  identify the correct set of dimensions by involving  other considerations.

Character table is a compact way of describing all irreducible representations of a given group. It is a square $n_\calG\times n_\calG$ table, where columns correspond to different conjugacy classes, and rows are labeled by different irreducible representations. The entry at an intersection of $i$-th row and $j$-th column is given by the value of the character for the $i$-th representation of the group elements from the $j$-th conjugacy class.

Using the characters table of a given group, one can easily find multiplicities $a_i$ in the decomposition of 
a representation $D$ into irreducible representations, Eq.~(\ref{Eq:Gtoirrep}).  Provided, characters of the representation $D$, $ \chi(\calC_k)$, are known, we can find $a_i$ as  
\begin{equation} \label{Eq:achi}
  a_i
  =
  \frac{1}{h_\calG} \sum_{k=1}^{n_\calG}  N_k\, \chi^{(i)*}(\calC_k)\, \chi(\calC_k) ,
\end{equation}
where $h_\calG$ is the number of elements in $\calG$, and $N_k$ is the number of elements in the corresponding conjugacy class.

If representation $D$ is obtained as the tensor product of two representations, let us say, $E$ and $F$, $D = E\times F$, the  characters of $D$, $\chi(\calC_k)  \equiv  \chi^{E\times F}(\calC_k)$, can be obtained as a product of characters for representations $E$ and $F$,
\begin{equation} \label{Eq:chiEF}
  \chi^{E\times F}(\calC_k) = \chi^E(\calC_k) \chi^F(\calC_k).
\end{equation}
After this, one can easily apply formula~(\ref{Eq:achi}) to find the decomposition of the $E\times F$ into irreducible representations.

\begin{table}
\begin{center}
\begin{tabular}{l|ccccc}
\hline\hline
Rep.&   $E$ &   $\calC_2$   &   $\calC_4$  &   $\calC_{xy}$  &   $\sigma_{uv}$  \\ \hline
$A_1$   &   1   &   1       &       1   &       1           &       1                       \\ 
$A_2$   &   1   &   1       &       1   &      -1           &      -1                       \\ 
$B_1$   &   1   &   1       &      -1   &       1           &      -1                       \\ 
$B_1$   &   1   &   1       &      -1   &      -1           &       1                       \\ 
$E_1$   &   2   &   -2      &       0   &       0           &       0                       \\ \hline\hline
\end{tabular}
\end{center}
\caption{\label{Tab:C4v} Irreducible representations of $C_{4v}$ and their characters.}
\end{table}

\subsection{Group of square lattice and its representations \label{AA:square}}
Here we illustrate how the facts summarized above  may be used to classify representations of the point symmetry group of square, $C_{4v}$ and its extension, $C_{4v}'$.

\begin{table*}
\begin{center}
\begin{tabular}{l c c c c c c ccc cccc}
\hline\hline
Conj. class    & $E$ & $\calC_{\bt}$      & $\calC_{\bt\bt}$         & $\calC_2$          &   $\calC_{2\bt}$ &$\calC_{2\bt\bt}$  & $\calC_{4}$                        \\
 $N_\calC$  & 1         &   2                & 1                        &  $1      $         &     2            &   1                & 4                                                  \\
Members     & $\1$      &$\SO{\1}{\ba_{1,2}}$& $\SO{\1}{\ba_{3}}$     &  $\SO{\Rot_{\pi}}{0}$       & $\SO{\Rot_{\pi}}{\ba_{1,2}}$    &$\SO{\Rot_{\pi}}{\ba_3}$  &$\SO{\Rot_{\pi/2}^{1,3}}{0,\ba_3}$   \\
\hline\hline
Conj. class    & $\calC_{4\bt}$       &$\calC_{xy}$                 & $\calC_{xy\bt_1}$            & $\calC_{xy\bt_2}$                   & $\calC_{xy\bt\bt}$                & $\calC_{uv}$            & $\calC_{uv\bt}$             \\
$N_\calC$   &4 &  2                          &   2                          & 2                                   &   2                            &   4                    & 4           \\
\mr{Members}     & \mr{$\SO{\Rot_{\pi/2}^{1,3}}{\ba_{1,2}}$}  & \mr{$\SO{\Ref_{x,y}}{0}$}      &$\SO{\Ref_{x}}{\ba_{1}}$& $\SO{\Ref_{x}}{\ba_{2}}$      &\mr{$\SO{\Ref_{x,y}}{\ba_{3}}$}  &\mr{$\SO{\Ref_{u,v}}{0,\ba_3}$} &\mr{$\SO{\Ref_{u,v}}{\ba_{1,2}}$} 
\\
       &          &                   &$\SO{\Ref_{y}}{\ba_{2}}$&$\SO{\Ref_{y}}{\ba_{1}}$           &                               &                             &                               \\
\hline\hline
\end{tabular}
\caption{\label{Tab:C'4vConj} Labeling of conjugacy classes of group $C'_{4v}$.  Below each label, number of group elements, $N_\calC$, belonging to a given conjugacy class, as well as explicit form of these elements in Seitz notations are given. Vector $\ba_3$ is a short-hand notation for the sum of lattice vectors, $\ba_3=\ba_1+\ba_2$.}
\end{center}
\end{table*}

\subsubsection*{Point group of square $C_{4v}$ and its representations \label{AAA:C4v}}

We start with reviewing properties and representations of the point symmetry group of square, $C_{4v}$. This is the group of all symmetry operations, which leave square invariant. It can be generated by rotations for $\pi/2$ around the center of the square, $\Rot_{\pi/2}$ and a reflection of $x$-axis, $\Ref_x$.~\cite{GroupTheory} In total the group $C_{4v}$ has $h_{C_{4v}}=8$ elements. In addition to rotations for angles multiple of $\pi/2$, these include reflections around $x$ and $y$-axes, as well as $\Ref_{u,v}$, standing for reflections relative to the planes containing vectors $\bx\pm\by$, $\Ref_{u,v}=\Ref_{x,y}\Rot_{\pi/2}$.

These elements can be split into total of $n_{C_{4v}}=5$ conjugacy classes. There are two conjugacy classes consisting of only one group element: trivial $E=\{\1\}$, and $\calC_2$ consisting of rotation for $\pi$, $\calC_2=\{\Rot_{\pi}\}$. Each of the remaining three classes consists of two elements: $\calC_4=\{\Rot_{\pi/2},\Rot_{3\pi/2}\}$, $\calC_{xy}=\{\Ref_{x},\Ref_{y}\}$, and  $\calC_{uv}=\{\Ref_{u},\Ref_{v}\}$. Correspondingly, group $C_{4v}$ has five irreducible representations. Using Eq.~(\ref{Eq:dim-irreps}) we find that four of irreducible representations are one-dimensional and one is a two-dimensional.  Characters of these irreducible representations are listed in Table~\ref{Tab:C4v}. One-dimensional representations are fully specified by their list of characters. Whereas two dimensional representation $E_1$ corresponds to a transformation of a vector. If we denote the basis of $E_1$ as $(\bx, \by)$, action of the group generators  becomes
\begin{subequations}\label{Eq:bxby}
\begin{eqnarray} 
  \Rot_{\pi/2}:& \quad \bx \to \by,& \quad \by \to -\bx,\\
  \Ref_{x}:& \quad \bx \to -\bx,& \quad \by \to \by.
\end{eqnarray}
\end{subequations}

Using Table~\ref{Tab:C4v} we can easily find decomposition of the Kronecker product of $E_1\times E_1$ into irreducible representations.~\cite{GroupTheory} Only two non-zero characters of $E_1\times E_1$ are $\chi^{E_1\times E_1} (E) =\chi^{E_1\times E_1} (\calC_2) = 4$. Now, using Eq.~(\ref{Eq:achi}) we can find that first four representations in Table~\ref{Tab:C4v} are contained once within $E_1\times E_1$: corresponding multiplicities are all equal to one, $a_i = 1/8\cdot(4+4)=1$. Whereas for $E_1$, we find corresponding $a$ to be zero. This may be summarized as 
\begin{equation} \label{Eq:E1E1app}
  E_1\times E_1
  =
  A_1\oplus A_2\oplus B_1\oplus B_2
  .
\end{equation}
Although formula~(\ref{Eq:achi}) gives us information about representations contained within $E_1\times E_1$, it does not give explicit expression for basis of these irreducible representations. In the present case the explicit form of the basis may be easily guessed from physical arguments.   Basis of each $E_1$ in the product can be written as a two components of a vector, with the action of generators specified in Eq.~(\ref{Eq:bxby}). Having components of two vectors $(q_x,q_y)$ and $(u_x, u_y)$, one can easily guess that the quantity, invariant under all symmetries is the scalar product. Thus, ${\bq}\cdot {\bm u}=q_x u_x + q_y u_y$ is a basis of $A_1$ component, contained in Eq.~(\ref{Eq:E1E1app}). Basis for $A_2$ is also easy to guess, as it has to change sign under any reflections. Thus,  it is given by the vector product, ${\bq}\times {\bm u}=q_x u_y - q_y u_x$. Finally, one can check that remaining combinations $q_x u_x - q_y u_y$ and $q_x u_y + q_y u_x$ realize the basis for $B_1$ and $B_2$ irreducible representations. This leads us to the Eq.~(\ref{Eq:E1E1bases}) in the main text, which summarizes the above results.

\subsubsection*{Group $C'_{4v}$ and its representations \label{AAA:C4v'}}

From the  group $C_{4v}$ we move to the group $C'_{4v}=G_\square/G_{2t}$, which is  the factor group of the space group of square lattice $G_\square$ over the group of translations for two unit cell vectors $G_{2t}$. In other words, group $C'_{4v}$ is defined as group $C_{4v}$ with added translation operations $t_{{\bm a}_1}$ and $t_{{\bm a}_2}$. In order to specify this group we will use Seitz operators $\SO{R}{\bt}$ defined as
\begin{equation}
  \SO{R}{\bt}\cdot \br = R\cdot\br + \bt.
\end{equation}
The group
\begin{equation}
  G_{t}
  =
  \{\SO{\1}{0}, \SO{\1}{\ba_1}, \SO{\1}{\ba_2}, \SO{\1}{\ba_1+\ba_2} \}
\end{equation}
is the subgroup of $C'_{4v}$ and it contains $h_{G_t}=4$ elements. The group $C'_{4v}$ has $h_{C'_{4v}}=h_{G_t}\cdot h_{4v}=32$ elements. It has $n_{C'_{4v}}=14$ conjugacy classes, which are listed in Table~\ref{Tab:C'4vConj}.
\begin{table*}
\begin{center}
\begin{tabular}{l | c c c c c c ccc ccccc }
\hline\hline
Rep.     & $E$ & $\calC_{\bt}$      & $\calC_{\bt\bt}$         & $\calC_2$          &   $\calC_{2\bt}$ &$\calC_{2\bt\bt}$  & $\calC_{4}$                 & $\calC_{4\bt}$                  &  $\calC_{xy}$                 & $\calC_{xy\bt_1}$ & $\calC_{xy\bt_2}$                      & $\calC_{xy\bt\bt}$                & $\calC_{uv}$            & $\calC_{uv\bt}$         \\
\hline
$A_1$ &1  & 1 & 1 & 1 & 1 & 1 & 1 & 1 & 1 & 1 & 1 & 1 & 1 & 1  \\
$A_2$ &1  & 1 & 1 & 1 & 1 & 1 & 1 & 1 & -1& -1& -1& -1& -1& -1\\
$B_1$ & 1 & 1 & 1 & 1 & 1 & 1 & -1& -1& 1 & 1 & 1 & 1 & -1& -1\\
$B_2$ & 1 & 1 & 1 & 1 & 1 & 1 & -1& -1& -1& -1& -1& -1& 1 & 1 \\
$A_1'$ & 1 & -1& 1 & 1 & -1& 1 & 1 & -1& 1 & -1& -1& 1 & 1 & -1 \\
$A_2'$ & 1 & -1& 1 & 1 & -1& 1 & 1 & -1& -1& 1 & 1 & -1& -1& 1 \\
$B_1'$ & 1 & -1& 1 & 1 & -1& 1 & -1& 1 & 1 & -1& -1& 1 & -1& 1 \\
$B_2'$ & 1 & -1& 1 & 1 & -1& 1 & -1& 1 & -1& 1 & 1 & -1& 1 & -1\\
$E_1$ & 2 & 2 & 2 & -2& -2& -2& 0 & 0 & 0 & 0 & 0 & 0 & 0 & 0  \\
$E_1'$ & 2 & -2& 2 & -2& 2 & -2& 0 & 0 & 0 & 0 & 0 & 0 & 0 & 0 \\
$E_2$ & 2 & 0 & -2& 2 & 0 & -2& 0 & 0 & -2& 0 & 0 & 2 & 0 & 0 \\
$\En_3$ & 2 & 0 & -2& 2 & 0 & -2& 0 & 0 & 2 & 0 & 0 & -2& 0 & 0 \\
$\En_4$ & 2 & 0 & -2& -2& 0 & 2 & 0 & 0 & 0 & 2 & -2& 0 & 0 & 0 \\
$\En_5$ & 2 & 0 & -2& -2& 0 & 2 & 0 & 0 & 0 & -2& 2 & 0 & 0 & 0 \\
\hline\hline
\end{tabular}
\caption{\label{Tab:C'4virreps} Irreducible representations of $C'_{4v}$ and their characters. The first eight representations are one-dimensional, the remaining six representations are two-dimensional.}
\end{center}
\end{table*}

Representations of $C'_{4v}$ can be worked out using the fact that it has a subgroup $G_t$. Consequently we can easily obtain five irreducible representations,  one-dimensional $A_{1,2}$ and $B_{1,2}$ along with two-dimensional $E_1$,  from corresponding irreducible representations of~$C_{4v}$. For this we simply assume the action of translations to be  trivial. Assuming that translation result in multiplying basis elements by minus one, we find additional four one-dimensional irreducible representations, denoted as $A'_{1,2}$ and $B'_{1,2}$ to emphasize that these are an extension of corresponding representations from~$C_{4v}$. Analogous extension of $E_1$ is denoted as $E_1'$. Remaining four two-dimensional irreducible representations can be found explicitly using $SU(4)$ generators given by $\{\mu^i,\tau^i,\mu^i\tau^j\}$ as a basis. Action of translations for representations $E_2 \ldots E_5$ can be written as
\begin{equation} \label{Eq:txty}
  \Tra_{x,y}:\quad  \bx \to \mp \bx,\quad  \by \to \pm \by,
\end{equation}
 so that $\Tra_x \Tra_y  =-\1$. However, the transformation of basis under rotation and reflection are realized differently for each of these representations. For representations $E_2$ and $E_3$ we have
\begin{subequations}\label{Eq:bxbyE23}
\begin{eqnarray} 
  \Rot_{\pi/2}:& \quad \bx \to \by,& \quad \by \to \bx,\\
  \Ref_{x}:& \quad \bx \to \mp \bx,& \quad \by \to \mp \by,
\end{eqnarray}
with the minus (plus) sign corresponding to $E_2$ ($E_3$). For $E_4$~($E_5$) we get: 
\end{subequations}
\begin{subequations}\label{Eq:bxbyE45}
\begin{eqnarray} 
  \Rot_{\pi/2}:& \quad \bx \to \mp\by,& \quad \by \to \pm \bx,\\
  \Ref_{x}:& \quad \bx \to \mp \bx,& \quad \by \to \pm\by.
\end{eqnarray}
\end{subequations}
The  character table may be easily calculated from here, and it is summarized in Table~\ref{Tab:C'4virreps}.

From characters we determine the  decomposition of different representations of $C'_{4v}$ on fermion bilinears into irreducible representations. Indeed, basis in the space of all possible fermion bilinears that are singlets in spin sector can be constructed using $SU(4)$ generators $\{\mu^i,\tau^i,\mu^i\tau^j\}$. Therefore this problem is equivalent to reducing adjoint representation of $C'_{4v}$ on sixteen $4\times 4$ matrices $\{\1,\mu^i,\tau^i,\mu^i\tau^j\}$. Representation is fully specified by the action of generators. For the cases of the \piFs\ phase these are given by Eqs.~(\ref{Eq:Rx})-(\ref{Eq:TCPsi}). Whereas for the \sFs\ phase, the reader is referred to Ref.~\onlinecite{Hermele-sF}.~(Note, that there  $\Ref_x$ is defined as a reflection with respect to the edge of square,  whereas in our conventions reflection plane goes through the center of plaquette.  Therefore $\Ref_x$ from Ref.~\onlinecite{Hermele-sF} coincides with $\Tra_x\Ref_x$ in our notations.) Calculating characters and applying Eq.~(\ref{Eq:achi}), we find 
\begin{multline} \label{Eq:Gdecomp-piFA}
  G^\text{\piFs}_{\psi^\dagger \psi}
  =
  A_1+A_2+\Bn'_1+\Bn'_2+E_1+\En'_1
  \\
  +E_2+\En_3 +\En_4+\En_5,
\end{multline}
for the \piFs\ phase, where basis in terms of products of Pauli matrices for each irreducible componentis listed in Table~\ref{Tab:C'4vpiF} in the main text. We also obtained the same expressions for bases of different representations using the notations from Ref.~\onlinecite{Hermele-piF}. Analogously, for the \piFs\ phase we have:
\begin{multline} \label{Eq:Gdecomp-sFA}
   G^\text{\sFs}_{\psi^\dagger \psi}
  =
  A_2+B_1+\Bn'_1+\An'_2+2\En'_1+2\En_3+\En_4+\En_5,
\end{multline}
with details on the basis listed in Table~\ref{Tab:C'4vsF}. From here  we immediately recover result of Refs.~\onlinecite{Hermele-piF,Hermele-sF}  that no invariant fermion bilinear terms exist in \piFs\ and \sFs\ phases. Indeed, $G^\text{sF}$ does not contain trivial representation $A_1$. Whereas, even though $G^{\pi\text{F}}$ contains $A_1$, as one can see from Table~\ref{Tab:C'4vpiF} it is not invariant under time reversal, $\calT$, nor under charge conjugation, $\calC$.

\begin{table}
\begin{tabular}{c|rrrrrr} 
\hline\hline
Rep & $E$ & $\calC_2$ & $\calC_3$ & $\calC_6$ & $\calC_a$ &
$\calC_{a'}$
\\ \hline $A_1$ & 1 & 1 & 1 & 1 & 1 & 1 \\  
$A_2$ & 1 & 1 & 1 & 1 & $-1$ & $-1$ \\ 
$B_2$ & 1 & $-1$ & 1 & $-1$ & $1$ & $-1$ \\ 
$B_1$ & 1 & $-1$ & 1 & $-1$ & $-1$ & $1$ \\
$E_1$ & 2 & $-2$ & $-1$ & $1$ & 0 & 0 \\ 
$E_2$ & 2 & 2 & $-1$ & $-1$ & 0 & 0 \\ \hline \hline\end{tabular}
\caption{Irreducible representations of the group  $C_{6v}$ and their characters.\label{Tab:C6v}}
\end{table}

\subsection{Group of honeycomb and kagome lattices}
Since an extensive details for kagome and honeycomb lattices are available in the literature,~\cite{Hermele-k,Basko}  we  only briefly summarize the basic facts for the symmetry group of the hexagon $C_{6v}$ and its extension for the \piFk\ phase. More details for the honeycomb lattice can be found in Ref.~\onlinecite{Basko}. 
\subsubsection*{Point group of hexagon}

For kagome and honeycomb lattices the relevant point group is that of a hexagon, denoted as $C_{6v}$. It has $h_{C_{6v}}=12 $ elements and can be generated by the rotation $\Rot_{\pi/3}$ and the reflection of $y$-axis, $\Ref_y$. It has six different conjugacy classes and six irreducible representations, of which four are one-dimensional, and remaining are two-dimensional. Using characters of $C_{6v}$ shown in Table~\ref{Tab:C6v}, we can write product of $E_1\times E_1$~as 
\be
  E_1\times E_1
  =
  A_1\oplus A_2 \oplus E_2.
\ee

\subsubsection*{$C_{6v}'$ for kagome lattice}
Anzats  for the algebraic spin liquid on Kagome lattice has a larger unit cell than the case without any fluxes. Thus, to classify fermionic bilinears,  we again have to consider enlarged group, $C_{6v}'$, which is the $C_{6v}$ with added translations for primitive lattice vectors $\ba_1$ and $\ba_2$. 

The group $C_{6v}'$~(or, $G_{s2}$ in notations of Ref.~\onlinecite{Hermele-k}) has been studied extensively and its conjugacy classes along with characters are listed in Tables III and IV  in Ref.~\onlinecite{Hermele-k}. Using this information, we may find the decomposition of the representation on bilinears as in Eq.~(\ref{Eq:C6vG}) with bases of corresponding irreducible components listed in Table~\ref{Tab:C6vsFk}. 

In the next order, we have to decompose the $E_1\times G^\text{\piFk}_{\psi^\dagger\psi}$ into irreducible representations. This leads us to Eq.~(\ref{Eq:E1GpiFk}) in the main text, where components $A_1$, $A_2$ and $E_2$, which are of interest for us originate from the tensor product of $E_1$ with another $E_1$, contained within Eq.~(\ref{Eq:C6vG}). This readily allows us to find the basis for these representations. 

\section{Calculation of the polarization operator \label{App:ImPi}}

In this appendix we calculate the imaginary part of the polarization bubble. We work using assumptions, specified in the main text.  In particular we restrict ourselves to the clean limit $ql\gg 1$ and assume the temperature to the the largest energy scale in the problem, $T\gg \vf q \gg \om$. Note, that we use explicit value of $N=2$ corresponding to spin. Since the polarization operator is proportional to $N$, one can easily restore the answer for the general case. 

We write the polarization operator, $\Pi^{(i)}$, corresponding to the interaction vertex $\tilde M^{(i)}(\hat \bq)$ as
\begin{widetext}
\be \label{Pi0t-test}
  \Im\Pi^{(i)}(i\om_n,\bq)
   =
   2 T
   \Im
  \int (d k)\sum_{m}
   \tr [ \tilde M^{(i)}_{\bk}(\hat \bq)  G_{\bk+\bq}(i\om_m+i\om_n)  \tilde M^{(i)}_{\bk+\bq}(\hat \bq)  G_{\bk}(i\om_m) ],
\ee
where $ (d k) = dk_x dk_y/(2\pi)^2$ is the short-hand notation for the momentum integration measure. 
The interaction vertex $ \tilde M^{(i)}(\hat \bq)$ as well as the Greens function are matrices in spinor space, and tracing in~(\ref{Pi0t-test}) goes over matrix indices. 
After analytical continuation, the imaginary part of the the matsubara sum of two Greens functions is written as, 
\be
 \Im
 \sum_{m}
 [G_{\bk+\bq}(i\om_m+i\om_n)]_{\alpha \beta} [G_{\bk}(i\om_m)]_{\gamma \delta}
 =
 \frac{1}{2\pi T}
 \int dz\,
 \left(\tanh\frac{z}{2T}-\tanh\frac{z+\om}{2T}\right)
 \Im [G^R_{\bk+\bq}(z+\om)]_{\alpha \beta} \Im [G^A_{\bk}(z)]_{\gamma \delta}.
\ee
where we restored internal indices. $G^{R,A}_{\bk}(z)$ stands for retarded (advanced) Greens function for real frequencies,
\begin{equation} \label{Eq:G-def}
  G^{R,A}_\bk(z,\bk)
  =
  \frac{z+\vf\btau\cdot \bk}{(z \pm i0)^2-\vf^2\bk^2}.
\end{equation}

In what follows, we will need the expression for the trace of numerators of two Greens functions with corresponding interaction vertices in Eq.~(\ref{Pi0t-test}). For the case of density coupling, defined in Eq~(\ref{Eq:M-dens}), we have $\tilde M^{(0)}_{\bk}(\hat \bq)=\1$, and the trace is evaluated as:
\begin{equation} \label{Eq:tracePi}
  \Ttr^{(0)}(z,\om,\bk,\bq)
  = 
  \tr[\1\cdot (z+\om+\vf\btau\cdot (\bk+\bq))\cdot \1 \cdot  (z+\vf\btau\cdot \bk)]
  =
  4 [(z+\om)z+ (\vf k)^2 +  \vf ^2k q \cos\theta],
\end{equation}
with $\theta$ being the angle between vectors $\bk$ and $\bq$. Note that there is an additional factor of two in~(\ref{Eq:tracePi}) from accounting for the (trivial) valley structure, whereas the factor of two originating from spin degrees of freedom is included in~(\ref{Pi0t-test}).
For the case of spinon-phonon coupling, arising in the next order of expansion in $\bk$, the $\tilde M^{(1)}_{\bk}(\hat \bq)$ is given by Eq.~(\ref{Eq:tildeMkxkx}) and the trace results in a cumbersome expression for $\Ttr^{(1)}(z,\om,\bk,\bq)$, which will be not listed here. 
Using expression for the imaginary part of Green's functions, we have:
\begin{multline}
\Im\Pi^{R(i)}(\om,\bq)
  =
  \pi
  \Im
  \int (dk)\,
  \int dz\,
  \left(\tanh\frac{z}{2T}-\tanh\frac{z+\om}{2T}\right)
  \Ttr^{(i)}(z,\om,\bk,\bq)
  \frac{1}{4\vf k}
  \left[
  \frac{\delta(z+\vf k)\delta(\om+\vf k'-\vf k)}{\vf k-\om}
  \right.
  \\
  \left.+
  \frac{\delta(z-\vf k)\delta(\om-\vf k'+\vf k)}{\vf k+\om}
  +
  \frac{\delta(z+\vf k)\delta(\om-\vf k'-\vf k)}{\vf k-\om}
  +
  \frac{\delta(z-\vf k)\delta(\om+\vf k'+\vf k)}{\vf k+\om}
  \right] .
\end{multline}
We drop last two terms in the square brackets since they correspond to interband transitions, and for $\om\ll \vf q$ they are not important. Also, we expand the difference between hyperbolic tangents, thus getting the derivative of the Fermi distribution function, denoted as $n'_F(z)$:
\begin{multline} \label{ImPi2}
\Im\Pi^{R(i)}(\om,\bq)
  =
  2 \pi \omega\int (dk)\,
  \int dz\,
   n'_F(z)
  \Ttr^{(i)}(z,\om,\bk,\bq) \frac{1}{4\vf k}
  \left[
  \frac{\delta(z+\vf k)\delta(\om+\vf k'-\vf k)}{\vf  k -\om}
  \right.
  \\
  \left.
  +
  \frac{\delta(z-\vf k)\delta(\om-\vf k'+\vf k)}{\vf  k +\om}
  \right] .
\end{multline}
Using $\delta$-functions, we may get rid of the integration over $z$. Integral over angle between vectors $\bk$ and $\bq$, denoted as $\theta$, can be done using the following expression: 
\be
  \int d\theta\,
  \delta(\pm \omega-\vf |\bk+\bq|+\vf k) F(\theta)
  =
 2 \theta(2k-q)  \frac{\vf k\pm \om }{v^2_0kq |\sin\theta^\pm_0|}
F(\theta^\pm_0),
\quad
\text{where}
\quad
\cos\theta^\pm_0 = \frac{\om^2}{2\vf ^2kq} \pm \frac{\om}{\vf  q} - \frac{q}{2k}.
\ee
This is valid in the limit when $\vf q\gg \om$. Note, that we included an extra factor $2$ to account for two possible  values of $\theta^+_0$ (and  $\theta^-_0$), assuming that the  $F(\theta^+_0) $ is the same for both solutions. The integration over $\theta$ in Eq.~(\ref{ImPi2}) yields:
\begin{equation} \label{ImPi3}
\Im\Pi^{R(i)}(\om,\bq)
  =
    \frac{\om }{4\pi v^3_0 q }
   \int_{{q}/{2}}^\infty dk\,  
  \left[
   n'_F(-\vf k)
  \frac{\Ttr^{(i)}(-\vf k,\om,\bk,\bq)|_{\theta = \theta_0^-} }
  {k|\sin\theta_0^-|}
  \right.
  \left.
  +
  n'_F(\vf  k)
  \frac{
  \Ttr^{(i)}(\vf  k,\om,\bk,\bq)|_{\theta = \theta_0^+}}
  {k|\sin\theta_0^+|}
  \right] .
\end{equation}

We notice, that expression in the square brackets in Eq.~(\ref{ImPi3}) does not vanish if we put $\omega$ to zero within it for the case of density coupling~[when $\Ttr(z,k,\theta)$  is given by Eq.~(\ref{Eq:tracePi})]. In this case, accounting for the fact that $n'_F(\vf  k)$ for the vanishing chemical potential is an even function, we have:
\begin{equation} \label{ImPi4-dens}
\Im\Pi^{R(0)}(\om,\bq)
  =
  \frac{4\om }{\pi \vf  q }
  \int_{{q}/{2}}^\infty dk\,   n'_F(\vf  k)  \sqrt{k^2 -(q/2)^2}
    =
 -\frac{4\om}{\pi \vf  q} T\log 2
.
\end{equation}
When calculating the integral we used the fact that the main contribution to the integral comes from $\vf k\sim T$, thus we may neglect by $q$ in the square root. This answer reproduces the results, available in the literature.~\cite{Dorey,VafekThesis,KhveschenkoPRB06,Guinea,DasSarmaPRB07} Recalling that this polarization operator is proportional to $N$, which was assumed to be $N=2$ for this calculation, we reproduce the imaginary part of the result listed in the main text, Eq.~(\ref{Eq:PiA-T}).

The calculation for the case of the next order coupling, $\tilde M^{(1)}_{\bk}(\hat \bq)$, requires more care. The answer depends on the direction of the phonon momentum, $\bq$. We define the $\phi$ to be an angle of $\bq$ relative to the $x$-axis, so that $\hat \bq = (\cos \phi, \sin\phi)$. Lengthy, but straightforward calculation gives for the polarization operator in this case
\begin{equation} \label{ImPi4-k}
\Im\Pi^{R(1)}(\om,\bq)
  =
   \frac{\om \sin^4 2 \phi }{\pi  q}
  \int_{q/2}^\infty dk\, n'_F(\vf k) \frac{ k^4  }{ \sqrt{k^2-(q/2)^2}}
    =
    -\frac{9  \zeta (3)  }{2\pi}  \frac{\om}{\vf^3 q} T^3 \sin ^4 2 \phi.
\end{equation}
Noteworthy, the answer is invariant under rotations of $\pi/2$, as one may expect for our case. The angular dependence of~(\ref{ImPi4-k}) is very anisotropic, in particular,  when $\bq$ points along $x$ or $y$-axes, the result vanish, indicating that the answer will be of higher order in $\omega$.
\end{widetext}

%

\end{document}